%% file: mag_gribov.tex
\documentstyle [art11,epsf]{article}
\newlength{\sze}
\newlength{\hspc}
\newlength{\vspc}
\def\s3		{\mathord{\sigma_{3}}}
\def\setl	{\left \{}
\def\setr	{\right \}}
\def\vmu	{\mathord{\widehat{\mu}}}
\def\dag	{\mathord{\dagger}}
\def\Tr		{\mathord{\hbox{Tr}}}
\def\mb#1	{\mbox{\boldmath $#1$}}
\def\fa		{\mathbin{\forall}}
\def\ng		{\mathord{n_{\hbox{\tiny G}}}}
\def\eff	{\mathord{\hbox{\tiny eff}}}
\def\diff	{\mathord{\hbox{\tiny Diff}}}
\title	{Gribov Copies in the Maximally Abelian Gauge and Confinement.}
\author	{A. Hart\thanks{e-mail: {\sl harta@thphys.ox.ac.uk}}\, 
and M. Teper\thanks{e-mail: {\sl teper@thphys.ox.ac.uk}}\,.
\\
{\small\sl Theoretical Physics, University of Oxford,
	1 Keble Road, Oxford, OX1 3NP, U.K.}}
\begin	{document}
\maketitle
\begin	{abstract}
\noindent
We fix $SU(2)$ lattice gauge fields to the Maximally
Abelian gauge in both three and four dimensions. 
We extract the corresponding $U(1)$ fields and monopole current
densities and calculate separately the confining string tensions
arising from these $U(1)$ fields and monopole `condensates'. 
We generate multiple Gribov copies and study how the $U(1)$ fields and
monopole distributions vary between these different copies.  
As expected, we find substantial variations in the number of
monopoles, their locations and in the values of the $U(1)$ field
strengths. The string tensions extracted from `extreme' Gribov 
copies also differ but this difference appears to be no more
than about 20\%.
We also directly compare the fields of different Gribov copies. 
We find that on the distance scales relevant to
confinement the $U(1)$ and monopole fluxes that disorder Wilson loops
are highly correlated between these different Gribov copies. 
All this suggests that while there is indeed a Gribov copy problem 
the resulting ambiguity is, in this gauge and for
the study of confinement, of limited importance.
\end	{abstract}

\vfill
Oxford Preprint Number: {\em OUTP--96--25--P}
\hfill
hep-lat/9606007

\newpage
\section	{Introduction.}
\label		{intro}

The idea of 't Hooft  
\cite{tHo}  
that confinement in non-Abelian gauge
theories might be associated with monopoles in suitable Abelian
projections of the fields, has been the subject of extensive
numerical investigation in recent years. 
This interest was provoked by the observation 
\cite{suzuki}
that in a particular gauge, the Maximally Abelian (MA) gauge 
\cite{MAG},
the string tension one obtains from the 
corresponding Abelian Wilson loops
appears to equal the full non-Abelian string tension
\cite{magrev}. 
Subsequently it was shown that the same is true for the string tension
that one obtains from the monopole currents (e.g. 
\cite{stack2}).

The MA gauge is, however, plagued by Gribov copies; each
non-Abelian field configuration has multiple gauge
copies along its gauge orbit which satisfy the MA gauge 
condition. 
The $U(1)$ fields and monopole currents that one extracts from these
different non-Abelian gauge copies are
gauge-variant and differ amongst the various Gribov copies
\cite{suzuki2,bali_Rmax}.  
Thus one expects that the $U(1)$ and monopole string tensions will
differ according to which Gribov copies one chooses to use in the
calculation. 
This has cast doubt on the significance of the comparison with the
non-Abelian string tension, given that those calculations relied on
choosing the Gribov copies at random.

What is the proper way to treat these Gribov copies is not known.
We focus in this paper on a useful preliminary question: 
what is the magnitude of the
variations in those properties of the $U(1)$ fields and monopole
currents which determine the string tension?
If the variation is large then no further progress is possible without
addressing the Gribov copy problem.
If, on the other hand, the variation is small then calculations which
ignore the Gribov copy problem should be reasonably reliable. 
We find that the latter appears to be the case in both 3 and 4
dimensions.

We carry out calculations simultaneously in 3 and 4 space-time
dimensions.
$SU(N)$ gauge theories appear to possess linear confinement in both
cases and in both cases this is a non-perturbative phenomenon.  
That confinement should be driven by monopoles is an argument that can
be made equally well in either case.

The contents of this paper are as follows. 
The procedure for fixing to the Maximally Abelian gauge is reviewed in
section~\ref{magfixing} and we indicate there why Gribov copies will
naturally arise. 
In section~\ref{gribcops} we show that Gribov copies are a
non-perturbative phenomenon: if the volume is so small that the
gauge field coupling is $\le O(1)$ on all length scales then,
effectively, no extra Gribov copies are generated.
On larger volumes such copies appear and we show how they differ in
relevant quantities such as the monopole density.  
In section~\ref{model} we outline a speculative picture of 
monopole confinement, in order to provide
a specific framework within which to discuss the
possible differences between Gribov copies. 
In section~\ref{gribcops_string} we present our calculations of the
string tensions for different subsets of Gribov copies. Although
our calculations are not unambiguous, they indicate a small
but systematic variation of the string tension with the choice
of Gribov copy ensemble.
Then in section~\ref{corrs} we compare directly the $U(1)$ fields and
monopole distributions of different Gribov copies, with a
focus on the long range structures that lead to confinement. 
We find that any such differences are remarkably small.
We finish with some conclusions in section~\ref{conclusion}.

The work in this paper is an outgrowth of a wider-ranging study of 
confinement in various Abelian gauges. We refer the reader to that
work \cite{tep3} for various technical
details, and for some of the calculations that are 
alluded to below.

\section	{Fixing to the Maximally Abelian Gauge.}
\label		{magfixing}

An $SU(2)$ lattice field configuration consists of a set of 
$SU(2)$ matrices
$\left\{ U_\mu(n) \right\}$
assigned to the links of a lattice of $V$ sites. To put this into the
MA gauge we find a local gauge transformation
$\left\{ g(n) \right\}$
which when applied to this field configuration
$$
U_\mu(n) \rightarrow U^{g}_\mu(n) = g(n) U_\mu(n) g^{\dag}(n+\vmu).
$$
maximises the gauge--dependent functional, $R$, of the
links:
$$
R = \frac{1}{V} \sum_{n} \Tr \left( X(n) \s3 \right) 
\hbox{\hspace{15ex} where \hspace{5ex}}
X(n) = \sum_{\mu > 0} U_\mu(n) \s3 U_{\mu}^{\dag} (n) 
$$
Maximising $R$ is equivalent to making $X(n)$ diagonal
(i.e. proportional to $\sigma_3$) for all $n$. 
It is also equivalent to maximising the sum over all links, $l$, of
the difference between the 11 and 12 components of the link matrices,
i.e. $\sum_{l} |u_{11}(l)|^2 - |u_{12}(l)|^2$.  
That is to say, it is the gauge in which the link matrices are made to
align, as closely as possible, along the $\sigma_3$ direction. 
Since matrices proportional to a given generator commute, this is 
called the Maximally Abelian gauge.

Once the field has been placed into this gauge, we write the link
matrices as the product of matter fields $c$ and Abelian fields,
represented as link angles,
$\setl \theta_{\mu} (n) \setr$,
via
$$
U^{g}_{\mu} (n)  =
\left(  \begin{array}{cc}
		c_{11} (n,\mu) & c_{12} (n,\mu) \\
	       -c_{12}^{*} (n,\mu) & c_{11} (n,\mu)
	\end{array}
\right)
\times
\left(  \begin{array}{cc}
		\exp i \theta_{\mu} (n) & 0 \\
		0 & \exp -i \theta_{\mu} (n)
	\end{array}
\right)
$$
where $c_{11}$ is real. 
The MA gauge fixing is incomplete and it is easy to see that the
remaining gauge transformations correspond to local Abelian gauge
transformations on the above Abelian fields. 
Abelian fields generically contain topological singularities which
correspond to magnetic monopoles, and these can be located using the usual
method of DeGrand and Toussaint \cite{DeGT}.

We can measure a $U(1)$ string tension by calculating large Wilson
loops with the above $U(1)$ fields.
We can also calculate the string tension produced by the monopole
currents, by iteratively solving 
\cite{tep3}
a set of dual Maxwell equations
\cite{stackvecpot}
to obtain a scalar potential in $D = 3$, or a
vector potential in $D = 4$.

Each $X(n)$ depends on the gauge transformations not only
at the site $n$ but also at neighbouring sites, so we cannot obtain an
immediate solution as we would if we had chosen to diagonalise an
operator such as the plaquette operator, $U_{\mu\nu}(n)$ (for some
values of $\mu, \nu$).
Instead we proceed iteratively using standard techniques
\cite{MAG}. 
If $x_{12}(n)$ is the 12 component of the matrix $X(n)$ then ideally
we should iterate until $x_{12}(n)=0$ for all $n$.  
In practice we iterate until, typically, $\vert x_{12}(n) \vert \le
10^{-7}, \fa n$.  
We have checked that with this criterion any gauge fixing systematic 
errors are far below our statistical errors.

\begin{figure}[p]
\begin{center}
\leavevmode
\epsfxsize=5.5in
\epsffile{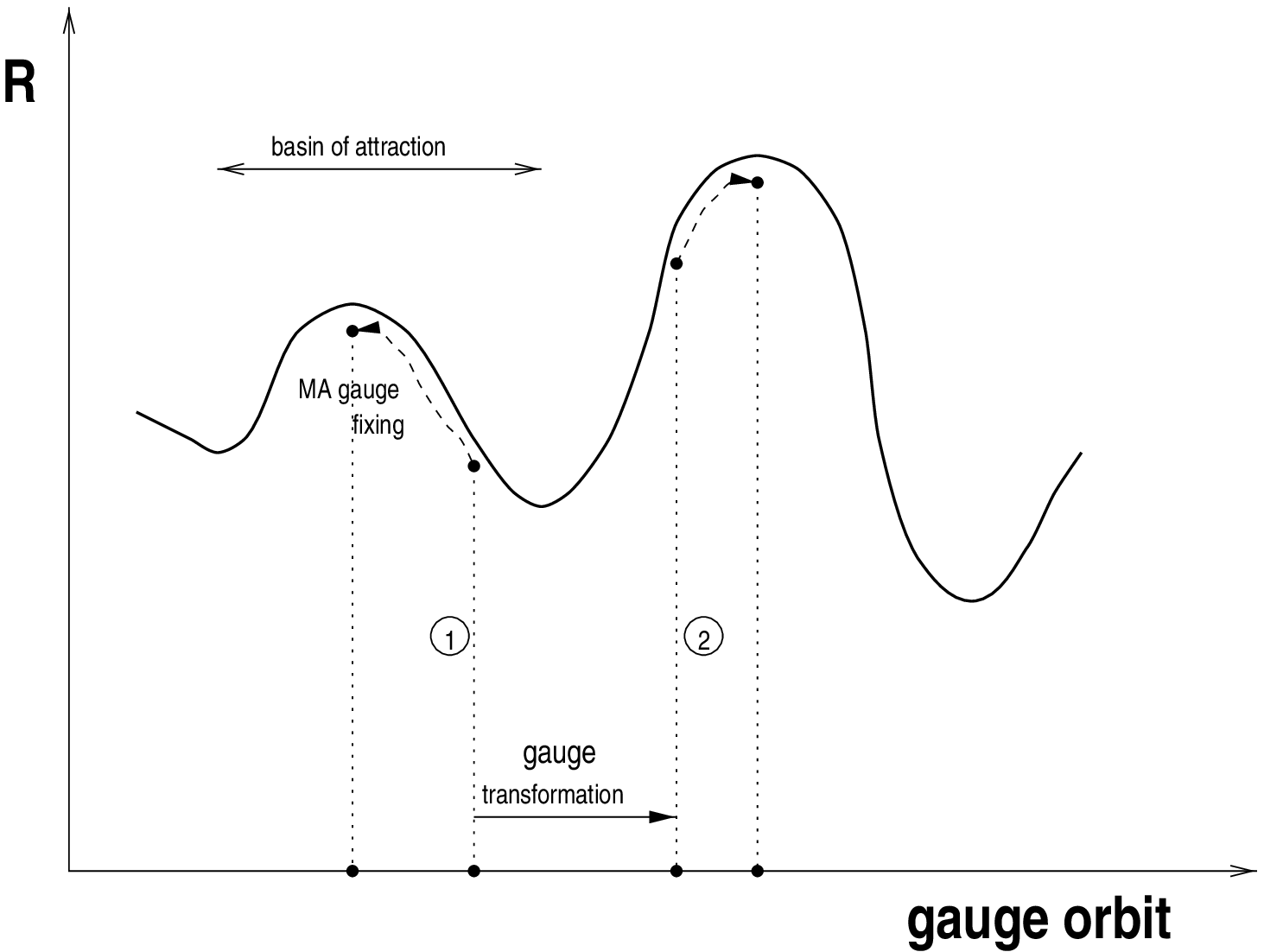}
\end{center}
\caption{A schematic diagram of the gauge orbit of a single $SU(2)$
configuration. The dotted lines denote the effect of iteratively
fixing to the MA gauge. This diagram is not designed to be accurate,
but to illustrate why two gauge transformations, 1 and 2, of a single
configuration may fix to different Gribov copies.}
\label{rform}
\end{figure}

The functional $R$ defined over the gauge orbit corresponding to some
particular field configuration has, in general, many maxima.
This is illustrated schematically in Figure~\ref{rform}.  
The gauge copy at each maximum satisfies the MA gauge condition, and
is a Gribov copy. 
(In this paper we shall refer to {\em all} these fields as Gribov copies,
rather than singling out one as the `original' and the remainder as
`copies'. In addition we disregard the obvious degeneracies 
that arise through the remnant $U(1)$ gauge invariance etc.) 
Although gauge invariant quantities are the same for all these copies,
the $U(1)$ fields defined above, together with their associated
magnetic current distributions, are only gauge invariant under the
remnant $U(1)$ sub-group, and so will in general differ. 
So, for example, if we were to calculate the $U(1)$ string tension
using always the field configuration corresponding to the global
maximum of $R$, then we might expect to get a different value than the
one we would get if we chose a Gribov copy at random
\cite{bali_Rmax}. 
Since the correct selection is not known, there is an obvious problem
of interpretation.

In a gauge where the operator $X(n)$ transforms purely in the adjoint
representation, such as the plaquette gauge mentioned above or the
Polyakov loop gauge \cite{suzukipoly}, the gauge fixing transformation can
be calculated exactly without iteration and there is just a single Gribov
copy.

\section	{Gribov Copies - a First Comparison.}
\label		{gribcops}

The numerical procedure for fixing to the MA gauge smoothly deforms a
field along its gauge orbit to a field at which $R$ is a local maximum.
Since the local deformations increase $R$, this procedure provides a
natural partition of the gauge orbit into subsets of gauge copies,
each of which subsets is associated with a specific Gribov copy.
In this picture, illustrated schematically in Figure~\ref{rform}, the
portion of the gauge orbit between neighbouring minima of $R$
forms a `basin of attraction' for the intervening maximum.
(Of course the exact details of 
the gauge fixing algorithm must have some effect on the boundaries 
of these subsets. We shall return to this question at the end of this
Section.)

For a given $SU(2)$ lattice field configuration, this partition can be
mapped out by applying an ensemble of random
gauge transformations to it.
Upon fixing to the MA gauge, each of these transformed fields will be
deformed into one of the Gribov copies.  
The fraction of these transformed fields that is associated with a
particular Gribov copy provides a natural measure of the fractional
volume of the gauge orbit associated with that Gribov copy.

If the field configuration is a typical field corresponding to the
coupling being weak at all length scales (as one would obtain in a
sufficiently small space-time volume) then one would expect to find 
only one Gribov copy, just as in perturbation theory. 
That is to say, as the coupling vanishes, the fraction of the gauge
orbit volume corresponding to one particular copy will go to
unity.  (As remarked previously we ignore the trivial degeneracies 
that arise because of the remnant $U(1)$ symmetry.) 
One would expect that this copy should be the one corresponding to the
absolute maximum of $R$.   

We test this numerically in $SU(2)$ (using the standard Wilson
plaquette action) in both $D=3$ and $D=4$. We shall 
denote the couplings in the two theories by $\beta_3 \equiv 
{4 \over {ag^2}}$ and $\beta_4 \equiv {4 \over {g^2}}$ 
respectively. All calculated quantities will be in lattice units.
We begin with the $D=3$ case.
We have taken an $8^3$ lattice and have generated 20 independent
$SU(2)$ field configurations at values of $\beta_3 \equiv 4/ag^2$
ranging from 4 to 12.
Over this range of $\beta_3$ the string tension, $\sigma$, varies from
$\surd\sigma = 0.41$ to 0.12 in lattice units \cite{tep}.  
In units of the physical length scale, $\xi \equiv 1/\surd\sigma$, the
lattice volume is small at $\beta_3 =12$ and reasonably large at
$\beta_3 =4$.  
That is to say, at $\beta_3=4$ the typical field configuration
contains long-distance non-perturbative physics, while at $\beta_3=12$
the typical field configuration will correspond to relatively weak
coupling at all accessible length scales.  
From each of the $SU(2)$ field configurations, we generate 
$N_{GT} = 50$ random gauge copies.  
This is intended to provide an approximation to the full gauge
orbit.  
Each of these gauge copies is then fixed to the MA gauge. 

We focus on three quantities, listed in Table~\ref{f_p_ng_3}.
The first is the fraction, $f$, of the gauge orbit that belongs to the
Gribov copy corresponding to the largest (observed) value of $R$.  
The second is the probability, $p$, that the Gribov copy with the
largest associated fraction of the gauge orbit is in fact the one for
which the value of $R$ is a maximum.  
This is intended as an estimate of the probability that the Gribov
copy corresponding to the absolute maximum of $R$ has the largest volume
of the gauge orbit.  
The third is the average number of different Gribov copies, $\ng$,
obtained when fixing the $N_{GT} = 50$ gauge copies to the MA gauge.
The quoted errors should obviously be taken as being no more than
indicative.

We see that at small $\beta_3$, where the lattice volume is large
enough to accommodate non-perturbative physics, we have many Gribov
copies and the Gribov copy with the largest value of $R$ plays a much
diminished r\^ole.   
(Although it is interesting to note that this r\^ole is still much
greater than that of other individual Gribov copies.)   
As $\beta_3$ grows the number of Gribov copies rapidly decreases and
the Gribov copy with the largest value of $R$ becomes the only
important one.  
The transition between these two regimes occurs for $\beta_3 \sim 7$
where the lattice size in physical units is $8/\xi \sim 1.7$.  
All this is consistent with the general expectations we outlined
earlier in this section.  
(We cannot, of course, prove that the Gribov copy we find with the
largest value of $R$ actually corresponds to the absolute maximum;
however given the pattern of our results it seems that this must be
the case except possibly at sufficiently small values of $\beta_3$.)

We have performed similar calculations in the $D=4$ theory, using
$N_{GT} = 100$ random gauge copies of each of 20 independent $SU(2)$
configurations on an $8^4$ lattice. 
We do this over the range $2.4 \leq
\beta_4 \leq 2.7$ where the volume changes from being reasonably
large in physical units to being very small.
The corresponding values of $f$, $p$, $\ng$ are listed in
Table~\ref{f_p_ng_4}, where the behaviour is clearly very similar to
that in the $D=3$ case.
The transition between the regime with many copies to the one
with few occurs at $\beta_4 \sim 2.5$ which corresponds to a lattice
size in physical units of $8/\xi \sim 1.5$ \cite{michtep}; again
similar to $D=3$.

\begin{figure}[p]
\begin{center}
\leavevmode
\setlength{\sze}{2.5in}
\setlength{\hspc}{0.5in}
\setlength{\vspc}{0.5in}

\hbox{
\epsfxsize=\sze
\epsffile{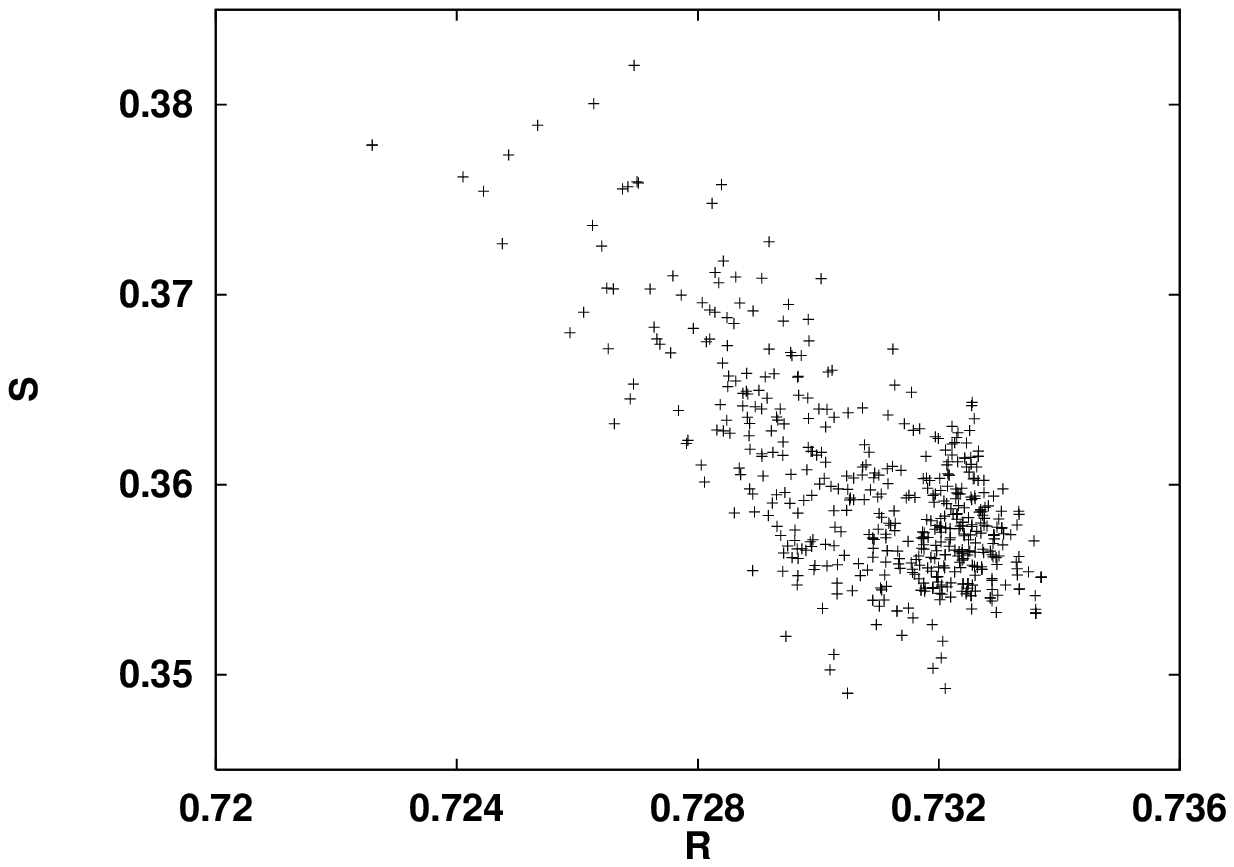}
\hspace{\hspc}
\epsfxsize=\sze
\epsffile{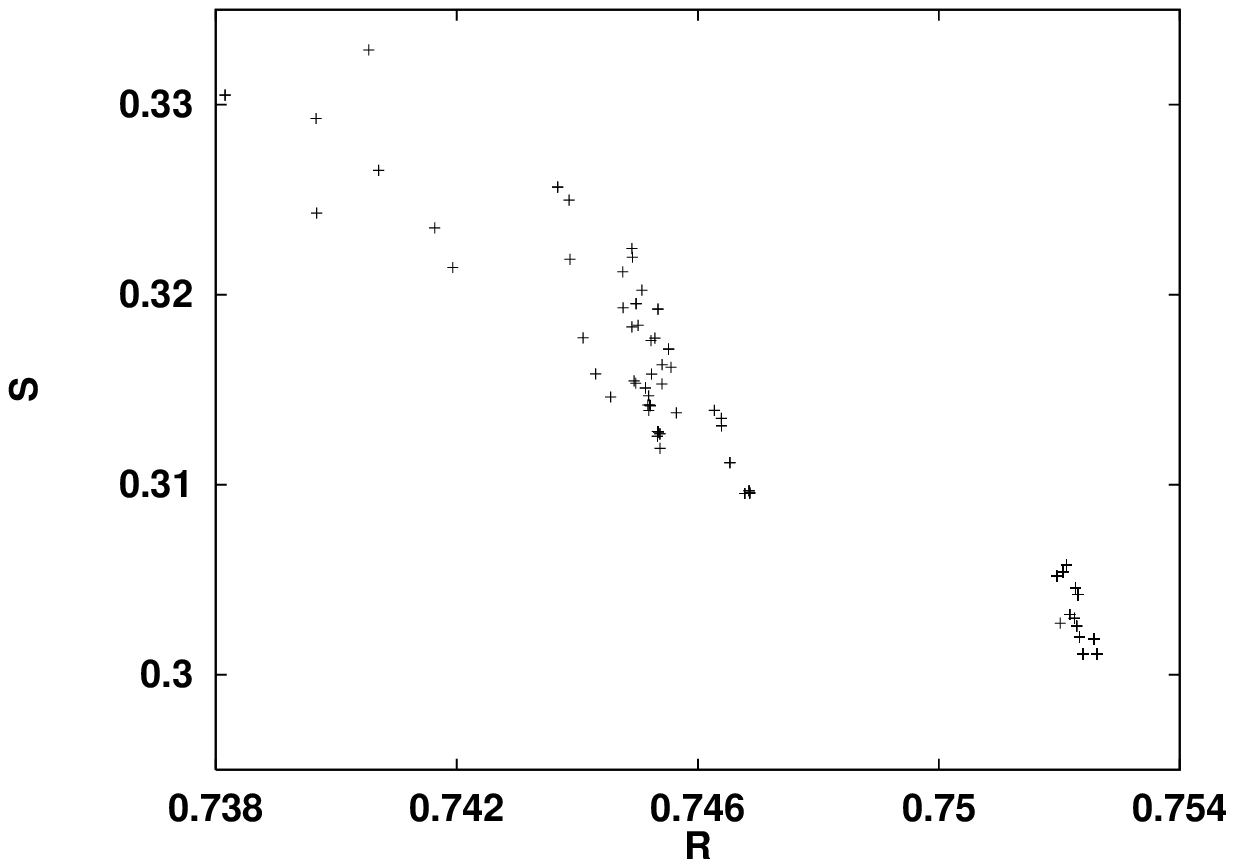}
}

\vspace{\vspc}
\hbox{
\epsfxsize=\sze
\epsffile{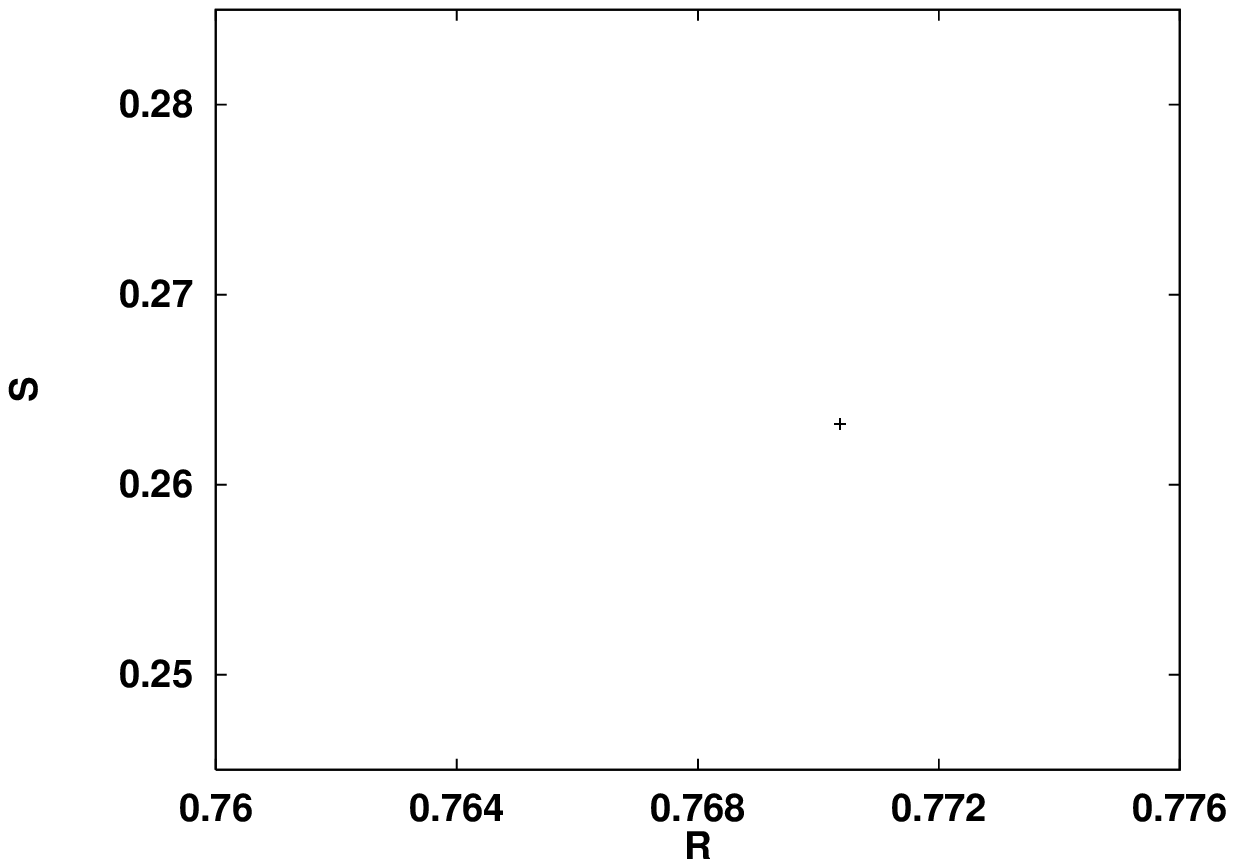}
\hspace{\hspc}
\epsfxsize=\sze
\epsffile{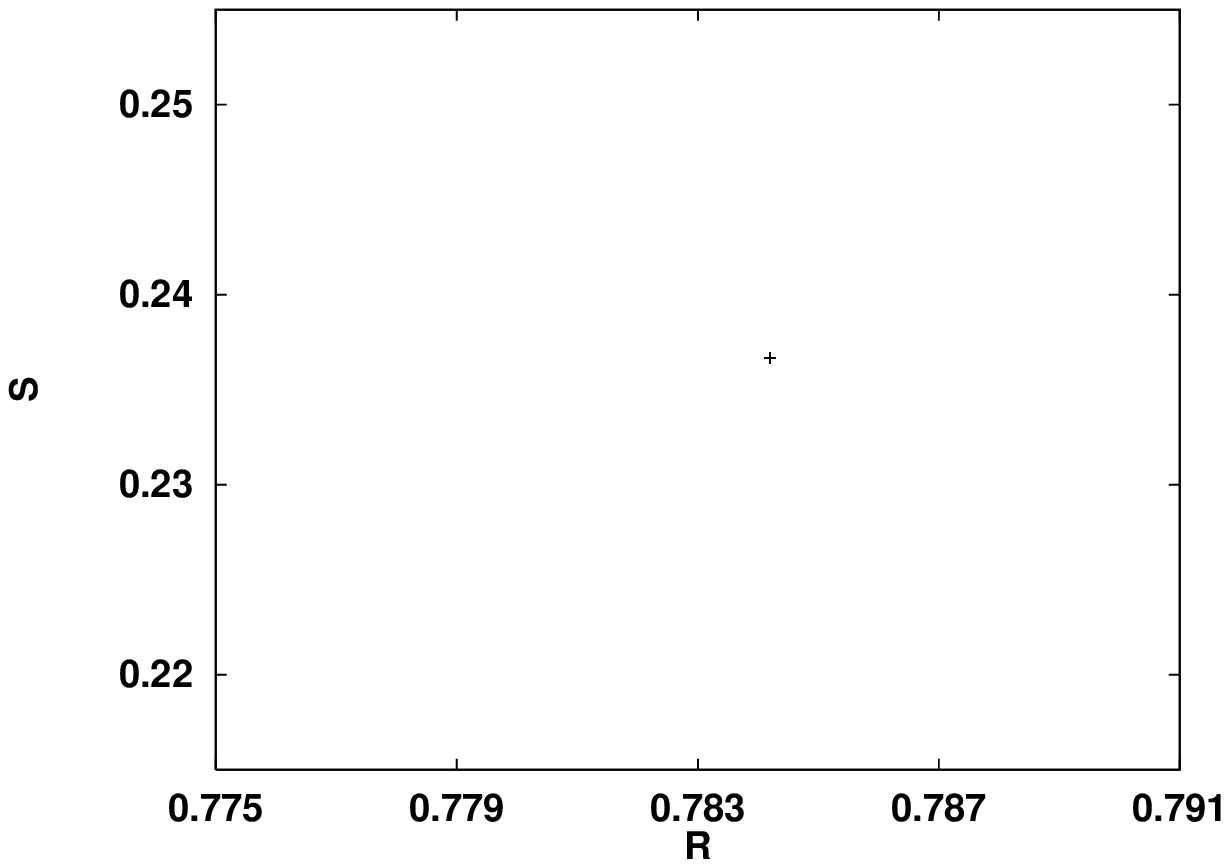}
}

\end{center}
\caption{Scatter plots of $S$ versus $R$ in $D=4$ at $\beta_{4} = 2.4,
2.5, 2.6, 2.7$ (reading horizontally). $N_{GT}=500$ for $\beta_{4} =
2.4, 2.5$, $N_{GT}=100$ for $\beta_{4} = 2.6, 2.7$.}
\label{srscatter}
\end{figure}

To see how the various Gribov copies differ with respect to the
$U(1)$ fields that we extract from them, we use a simple quantity that
is sensitive to the local fluctuations of the $U(1)$ fields, the
$U(1)$ plaquette action,
$$
S = \frac{1}{V} \sum_{p} \left( 1 - \cos \theta_p \right)
$$
where $\theta_p$ is the sum of the $U(1)$ link angles around 
the plaquette $p$.   
Note that this `action' has nothing to do with whatever is the
effective $U(1)$ action that describes the projected Abelian fields;
that is expected to be highly non--local \cite{magrev}.    
In Figure~\ref{srscatter} we display a scatter plot of $S$ versus $R$
for 500 MA gauge fixings obtained from a `typical' $8^4$
configuration at each of the values of $\beta_4$ indicated.   
Although there is a large scatter, there is an evident correlation
between larger values of $R$ and smaller values of $S$; that is,
Gribov copies corresponding to larger values of $R$ lead to $U(1)$
fields with weaker fluctuations
\cite{suzuki2}. This is also the case in $D=3$.

If we look at the monopole content of various Gribov copies, we find
that there are on average fewer monopoles in Gribov copies with larger
$R$, in both $D=3$ and $D=4$.
If confinement is driven by monopoles then, all other things being 
equal, the strength of the confining force will be proportional to the
strength of the monopole condensate. Our above observations would then 
imply that the string tension is smaller if one uses Gribov copies of 
larger $R$.   
This is a question we shall address more directly below.

Given that Gribov copies do differ it is natural to ask if there
is any convincing criterion for selecting which Gribov copies
should be used for extracting the $U(1)$ fields and monopole
distributions.   
 
One increasingly frequent suggestion is that the Gribov copy
corresponding to the absolute maximum of $R$ should be used.
This is partly motivated by an analysis of the corresponding
problem in the Landau gauge (see \cite{zwanz}
and references therein). A concrete example
is provided by Landau gauge calculations of the photon propagator 
in the Coulomb phase of U(1) lattice gauge theory. One finds 
\cite{Mitr1} 
that one gets an incorrect propagator unless one makes a
selection on the Gribov copies. If one selects the Gribov
copy corresponding to the absolute maximum (of the functional
appropriate to the Landau gauge) one indeed obtains the
correct perturbative propagator \cite{Mitr2}. 
This provides an argument for using the Gribov 
copy corresponding to the absolute maximum of $R$ in the case
where one wants to obtain quantities that are essentially
perturbative. (One should note however that even here this 
choice is not unique \cite{Mitr2}.)
It is however not at all clear that the
same criterion should apply when considering quantities
that are non-perturbative - or indeed
what lessons the Landau gauge has for the
Maximally Abelian gauge. 

Part of the difficulty in
motivating the choice of one particular Gribov copy is that
if we simply look at the various Gribov copies without any
theoretical prejudice then we find
that the values of $R$ corresponding to the various
Gribov copies, show very little variation.   
For example, on our $8^3$ lattice at $\beta_3=4$ the average value of
$R$ for all Gribov copies is 0.820 while the average value for those
copies with the maximum value of $R$ is 0.824.
On $12^4$ at $\beta_4=2.4$ the corresponding values are 0.7308 and
0.7320.
These are very small differences and it is hard to see what it is that
might pick out one maximum as the correct one to use. 

Finally we briefly return to the question raised at the 
beginning of this section. We described there how the
gauge orbit may be naturally partitioned into subsets, each of 
which is associated with a single Gribov copy. We then remarked
that our gauge-fixing algorithm would deform a particular
gauge copy into the associated Gribov copy, and so by
gauge-fixing an ensemble of randomly generated gauge copies
we would generate the Gribov copies with a probability
that was proportional to the volume of the associated subset
of the gauge orbit. This is clearly an idealisation and
a measure of our deviation from this idealised
picture can be provided by comparing the results of 
different ways of gauge fixing. In particular we can
see how the incorporation of over-relaxation alters the 
probability distribution of the resulting Gribov copies.
An added motivation for this comparison is that
intuitively the idealised picture holds best if we
make small gauge-fixing steps. However in order
to be reasonably efficient we are forced to incorporate 
the `larger' over-relaxation steps. One would
obviously like to know how much this biases our
resulting ensemble of Gribov copies. (Clearly, by
using steps that were large enough, one could imagine
an arbitrarily large bias of the distribution.)

We have taken an $8^3$ $SU(2)$ lattice field generated at 
$\beta_3=5$ and have generated 250 random gauge copies
from it. We have then gauge fixed in three different ways.
The first contained no over-relaxation steps. The second
contained 2 over-relaxation steps every 3 iterations
and corresponded to the value we typically used in our 
$D=3$ calculations.
The last had 4 over-relaxation steps every 5 iterations.
The corresponding average values of $R$ turned out to
be 0.8548, 0.8547, and 0.8551 respectively. The standard
deviations of the corresponding distributions were
0.0025, 0.0025, and 0.0020 respectively. Thus the overall
properties of the distributions of the Gribov copies
are very similar. Comparing the distributions in 
detail we find that the differences are located in the
long tail of Gribov copies with values of $R$ much
{\em below} the average. These Gribov copies have small
weights, individually, and so it is not clear whether
we are seeing a statistical or systematic effect.
The bulk of the Gribov copy distribution is 
almost identical and our results appear consistent 
with the differences being statistical. In particular
there appears to be no particular enhancement of the
number of Gribov copies with the larger values of $R$
when we alter the gauge-fixing procedure.

The example described in the previous paragraph appears to
be typical. It suggests that our idealised
picture does indeed make sense, except perhaps for the long
tail of Gribov copies with very small values of $R$.
\section	{A Simple Model.}
\label		{model}

To assess the significance of any differences that we observe
between different Gribov copies, it would be useful to have some
picture of the dynamics within which we can use our physical intuition.
Since this dynamics is not known, we shall sketch a picture which has
some plausibility even if it cannot be justified in detail (and indeed
is not completely consistent as it stands and may well be
incorrect).   

In this picture one supposes that at some length scale the gauge
fields produce a composite adjoint scalar field and that there is a
dynamical symmetry breaking on a longer distance scale analogous to 
the explicit breaking in the Georgi-Glashow model.    
Such a theory should possess `t Hooft-Polyakov monopoles. Outside
the cores of any monopoles the fields will be effectively $U(1)$, 
while within the cores they become fully $SU(2)$.  If these monopoles 
condense then they will generate a linearly rising potential
between static fundamental charges \cite{Polybook}.

If the volume of space outside the monopole cores is much
larger than the volume within the cores, so that the fields are
effectively $U(1)$ thoughout most of space, then we would
clearly expect the MA gauge to pick out this $U(1)$ field. That is 
to say if we go to the gauge where the composite scalar is
proportional to $\sigma_3$, then the gauge fields will be
essentially Abelian over most of space-time and we would
expect that this is what we would obtain by going to a gauge
that maximises the Abelian character of the $SU(2)$ fields.
Now if we consider a closed surface that encloses a monopole 
and if we locate this surface outside the monopole core,
i.e. in the region where the fields are Abelian, then there will be
a net magnetic flux out of that surface. Therefore if
we interpolate the $U(1)$ field within the region of
space-time occupied by the monopole core, this will
necessarily produce a Dirac magnetic monopole
somewhere in that region. 
Since within the core there is no physical $U(1)$ field, there
is no reason why the interpolated $U(1)$ field should not 
contain several (anti)monopoles in the region of space
occupied by the core. The only constraint is that the net 
magnetic charge within the core should equal the charge of the
't Hooft-Polyakov monopole. So any such extra 
monopoles will come in monopole-antimonopole pairs
whose separation is less than the core size.
Similarly even if there is a single monopole in the
core it need not be at the centre of the core. However
one may think of a shift in the position of a monopole as
equivalent to the addition of an appropriately positioned
monopole-antimonopole pair.    
Thus we would expect that
by going to the MA gauge not only would we obtain
the $U(1)$ field that is produced by the symmetry breaking,
but that we would also obtain a gas of Dirac monopoles whose
positions would be located within the cores of the
corresponding 't Hooft-Polyakov monopoles. The Dirac 
monopoles are of course unphysical, but they serve to
trace out the locations of the physical 't Hooft-Polyakov
monopoles. This gas of monopoles will in general include
a gas of monopole-antimonopole pairs whose separation is
no larger than the size of the core. That is to say,
a gas of magnetic dipoles. 

If the monopoles condense then we will get the same 
string tension whether we calculate it from the
't Hooft-Polyakov monopoles or from the $U(1)$
monopoles. The reason is that the extra Dirac 
monopoles are located in dipoles of a limited
spatial extent, and these can only contribute
to the shorter distance pieces of the potential.

In this picture we see that the density of $U(1)$ monopoles
does not of itself effect the string tension. Thus if
Gribov copies differ in that some contain extra
dipoles in the `t Hooft-Polyakov monopole cores, then
this will be harmless from the point of view of
confinement. Clearly we should compare the monopole
gases in different  Gribov copies and try to see whether
this is the case or not. We shall do this in a later
section.
 
In practice, of course, our non-Abelian theory has one 
overall scale and so there is no reason why the fraction 
of the volume inside the monopole cores should not be 
comparable to that outside.
Once a significant fraction of the space-time volume (i.e. that
within the monopole cores) does not contain a real $U(1)$
field, it is harder to see why the MA gauge fixing should pick out
the correct $U(1)$ field, or, if it does, why the Gribov copy for 
which the $SU(2)$ field is made as Abelian as possible {\em everywhere} 
should be the one that best corresponds to the dynamically generated
$U(1)$ field.
Indeed given the small observed differences in $R$ between the various
copies, it is quite possible that it is one of the copies with $R$
less than the absolute maximum that most faithfully maps out the
effective $U(1)$ fields and `t Hooft-Polyakov monopoles of this
picture.

\section	{Gribov Copies and the String Tension.}
\label		{gribcops_string}

We have seen in the previous section that just because 
Gribov copies differ in the apparent strengths of their magnetic 
condensates, this does not imply that they differ in their
confining properties. To address the latter issue, 
we shall compare the confining properties of those Gribov copies
that possess larger values of $R$ with those that possess smaller
values, and we shall compare both subsets with Gribov copies that are
simply chosen at random (as is usually done in calculations of the
Abelian string tension). 
There are many ways to perform such a selection and we have chosen the
following procedure. 

On each of the Monte Carlo generated $SU(2)$ fields we perform
$N_{GT}$ random gauge transformations.  
Each of these gauge copies is fixed to the MA gauge.  
This provides us with a number of Gribov copies that is $\le
N_{GT}$. 
(Since we work with volumes that are reasonably large, the number of
different copies will be close to $N_{GT}$.) 
From these Gribov copies we select the one with the smallest value of
$R$, the one with the largest value of $R$ and one at random. 
In this way we obtain from our ensemble of $SU(2)$ fields three
ensembles of $U(1)$ fields: one ensemble comes from copies with values
of $R$ that are smaller than average, one from copies with $R$ larger
than average and one from copies where $R$ is average (on the average).
We denote these ensembles as $\setl R_{min} \setr$, $\setl R_{max}
\setr$ and $\setl R_{av}\setr$ respectively.  
Within these three ensembles we calculate various quantities of
interest to determine in what ways Gribov
copies that have been so selected differ from each other. 
Clearly the larger $N_{GT}$, the greater the potential difference
between the ensembles.  
(Note that if we construct an ensemble from all the gauge copies with
equal weighting then this is equivalent to an ensemble where we choose
one at random, but should produce results that are more accurate.)

The quantities that we shall consider here are the $U(1)$ and 
monopole string tensions and, for comparison, the monopole density
and the average $U(1)$ plaquette.
The last two quantities are straightforward but the 
calculation of the string tension requires some explanation. 
The usual procedure would be as follows. 
On each of the configurations we calculate the values of the Wilson
loops, $W(r,t)$, both from the Abelian fields and from
the magnetic monopoles.
This provides us with averages on each of our three ensembles. 
Working within each ensemble separately, the quark potentials (in
lattice units) are extracted using the usual expression:
$$
V(r) = - \lim_{t \rightarrow \infty} \left[ \ln \left( 
\frac{\langle W(r,t+1) \rangle}{\langle W(r,t) \rangle} \right) \right] 
$$
The potential is then fitted with a sum of terms that are 
linear, Coulomb and constant in the distance $r$. 
The string tension, $\sigma$, is the coefficient of the linear term.

This basic method is not very efficient and in other contexts
it is usually improved upon in two ways. Firstly the fluctuations in
the time-like links can be reduced by a self-averaging
procedure. This makes most difference at smaller values of $\beta$.
For larger values of $\beta$ it can be improved upon enormously
by appropriately smearing the fields and by applying a variational 
criterion to the extraction of $V(r)$. 
This second technique also allows $\sigma$ to be calculated more
simply, and at least as accurately, from correlations of Polyakov
loops.

The first method of improvement cannot be applied here because
its details depend on the action and  the effective action
for the $U(1)$ fields obtained in the MA gauge 
is not known (and in any case is certain to be so non-local as
to render this method inapplicable).

The second improvement is only guaranteed to work within a
proper field theory; that is to say it depends on the existence of a
well-defined transfer matrix. 
Our $U(1)$ fields are derived from the $SU(2)$ fields by a very
non-local procedure and so there is no guarantee that the
corresponding ensembles have the desired properties.  
What we find, from calculations of correlation functions of 
smeared Polyakov loops in both 3 and 4 dimensions,
is that if such a transfer matrix does exist then it certainly 
does not have the positivity properties of the $SU(2)$ theory. 
In principle this need not be a problem as long as it does not affect
the large eigenvalues of the transfer matrix. 
(After all, most `improved' non-Abelian actions break positivity in
this sense.)
Unfortunately the loss of positivity has the practical defect of
making it difficult to apply the smearing+variational techniques that
have proved so powerful in extracting string tensions for non-Abelian
gauge theories. 

So if we want to calculate $\sigma$ we must rely on the inefficient
basic method outlined above. In the 4 dimensional case it turned
out that we were able to obtain accurate calculations of
the monopole string tension by this means. However we
were not able to do so for the $U(1)$ fields themselves.
This is both because they are relatively more noisy and also
because they contain larger sub-leading terms. 
(An alternative, and useful method is to calculate the
Wilson loops from `cooled' $U(1)$ fields \cite{tep3}.) 
In 3 dimensions there is the added complication that the 
Coulomb potential grows logarithmically.  
This is a particular problem in trying to obtain $V(r)$ and $\sigma$
from monopoles because here the Coulomb potential is not
screened. 
This additional long-range component makes the extraction of the
potential and string tension quite involved.  
For that reason we shall not present any results, in this paper, for
the $D=3$ monopole string tension.

To obtain an estimate of the confining properties of the $U(1)$ 
fields we use the oldest and simplest method of all, that of Creutz
ratios. 
We define Creutz ratios
$$
C(r,t) =  {{{\langle W(r,t)\rangle}\times{\langle W(r-1,t-1)\rangle}}
\over {{\langle W(r,t-1)\rangle}\times{\langle W(r-1,t)\rangle}}}
$$
and an effective string tension $\sigma_{\eff}(r)= -\ln C(r,r)$.
The string tension (in lattice units) is then obtained from    
$$
\sigma = \lim_{r \rightarrow \infty} \sigma_{\eff}(r).
$$ 
This definition can clearly be extended to make use of
Creutz ratios with $r \not= t$.

We start with the case of 3 dimensions. 
Taking advantage of the relative speed of 3 dimensional
computations, we perform calculations with a large number of gauge
copies, $N_{GT}=30$, so ensuring that our ensembles $\setl R_{min}
\setr$ and $\setl R_{max} \setr$ are really quite extreme. 
We do this on an ensemble of 400 independent $SU(2)$ gauge fields. 
In Table~\ref{cos_nm_3} we present the average $U(1)$ plaquette and
the average number of monopoles obtained on the three Gribov copy
ensembles, $\setl R_{max} \setr$, $\setl R_{min} \setr$ and $\setl
R_{av}\setr$, that we defined above.
The variation in the plaquettes is significant but small. 
On the other hand, the variation in the monopole densities is large. 
This is of concern since, all other things being equal,
the string tension will be proportional to the monopole density. 
Of course, as discussed in the previous section, it may well be 
that all other things are not equal. 
For example, if the difference in densities is due to
magnetic dipoles then there will be no effect on $\sigma$.
Whether there is an effect on $\sigma$ or not is the question we shall
now address.

In Table~\ref{sig_grib_3} we present the values of the effective
string tension for the three ensembles of Gribov copies.
We see that $\sigma_{\eff}$ appears to decrease towards its asymptotic
value, just as one would find if one carried out this calculation
directly with the $SU(2)$ fields. 
Moreover at any given value of $r$ we see that $\sigma_{\eff}(r)$
decreases as $R$ increases. 
What can we say about $\sigma$ itself?  
One would normally estimate $\sigma$ from $\sigma_{\eff}(r)$ at values
of $r$ where the latter had become independent of $r$ within
errors. 
Of course, this criterion will only work if the errors are
sufficiently small for the $r$-independence to be statistically
compelling. 
Clearly our data is rather marginal in this respect.
Nonetheless, using this criterion Table~\ref{sig_grib_3} suggests that
there is indeed a variation of $\sigma$ with the ensemble used.
If we take Gribov copies at random we find $\sigma \sim 0.89(3)$.  
Compared to this, $\sigma$ is reduced by about $10\%$ 
for the Gribov copies with the largest values of $R$
and increased by about $15\%$ for the Gribov copies with the smallest
values of $R$.
These differences are not negligible, but we find them remarkably
small given that we are effectively comparing the top
$\frac{1}{30}$th (in $R$) of the gauge orbit against the bottom
$\frac{1}{30}$th.
Certainly they are not so large as to render meaningless a comparison,
at least semi-quantitatively, with the full $SU(2)$ string tension,
which happens to be 0.0983(16) at $\beta_3 = 5$ \cite{tep}.

In the 4 dimensional case we have performed calculations on a $12^4$
lattice at $\beta_4=2.3$ and $\beta_4=2.4$. In the former case we have
produced ensembles of Gribov copies with $N_{GT}=10$ and in the
latter, where the statistical errors are smaller, with
$N_{GT}=5$. Because the values of $N_{GT}$ are much smaller
we expect the different $R$ ensembles
to be much less extreme than in the $D=3$ case.
In Table~\ref{cos_perim_4} we show how the average $U(1)$ plaquette
and the summed dual monopole current links
vary across these different ensembles for $\beta_4=2.4$.
The pattern of these results is similar to that obtained in $D=3$.

As stated earlier, we have been able to extract the $D=4$ 
monopole string tensions and these are listed in 
Table~\ref{sig_mon_4}. We see a significant variation
between the string tensions at $\beta_4=2.4$: the
string tension in the ensemble $R_{min}$ is  
$17\pm5\%$ greater than that in $R_{max}$, while that
in $R_{av}$ is about $10\pm5\%$ greater.
The $\beta_4=2.3$ values of $\sigma$ show a similar variation
with $R$.
For the $U(1)$ fields we follow the same procedure as in
$D=3$ and extract effective string tensions from Creutz
ratios. These are listed Table~\ref{sig_u1_4}. If we
compare the values of $\sigma_{\eff}(r)$ at, say, $r=3$
we observe a similar trend to that observed in the 
$D=3$ case.  

In all the above cases the trend is the same: the string tension
decreases as $R$ increases. Moreover it would appear from
these calculations that if we used the Gribov copy with the 
largest value of $R$ rather than choosing a copy at random
(as has usually been done in previous calculations) this
would reduce $\sigma$ by $O(10\%)$. 

We conclude from the above that while we obtain very different
monopole densities on ensembles of Gribov copies selected 
according to whether $R$ is large or small, the differences 
in the string tensions are much more modest. 

\section	{Gribov copies - a Direct Comparison.}
\label		{corrs}

So far we have compared the average properties of different subsets of
Gribov copies. 
The long-distance fluctuations can, however, be completely different
even if the average properties are the same. 
For example if we compare the $U(1)$ fields obtained from two
independent sets of $SU(2)$ gauge fields, the average properties will
be the same even though the long-distance fluctuations are completely
uncorrelated. 
Clearly to obtain a complete picture of how Gribov copies differ
we need to determine how correlated are the long-distance  
fluctuations of Gribov copies from the {\em same} $SU(2)$ gauge
field.

The `long-distance fluctuations' that are relevant to this study are
those that have to do with confinement.
This has to do with the area decay of large Wilson loops.
In a given $U(1)$ field configuration the value of a space-like
Wilson loop along a contour $\cal C$ is just a phase and can be
written as
$$
W({\cal C}) = e^{iB({\cal C})}
$$
where $B(\cal C)$ is the magnetic flux through a surface spanning that
contour.
(In Euclidean space-time, and at zero temperature, we can
confine our discussion to space-like loops with no loss of
generality. 
Note also that in 3 dimensions we follow convention and refer to the
fluxes as `magnetic', as though we were looking at static fields
within a time-slice of the 4 dimensional theory.)  
The expectation value of $W({\cal C})$ will depend on the distribution
of magnetic fluxes in the vacuum.  
In particular, the fact that the average value of $W({\cal C})$
decreases exponentially with the minimal area spanned by the contour
$\cal C$, tells us that the flux $B$ is effectively the sum of a
number of elementary fluxes which are mutually independent and
sufficiently localised so that the number of such fluxes passing
through the contour $\cal C$ is proportional to the minimal area
spanned by $\cal C$. 

One possible source of such fluxes is a gas of magnetic
monopoles. 
Such a gas has to have particular characteristics; for example, if the
(anti)monopoles pair off into dipoles then this will not disorder
large Wilson loops sufficiently to make them decay exponentially with
the area of the loop.
A screened plasma would, on the other hand, confine. 
However the locations of the monopoles are not really the relevant
degrees of freedom. 
For example if one has a confining monopole gas and shifts the
locations of the (anti)monopoles by a fixed distance in random
directions then this new gas is equivalent to the old gas plus a gas
of randomly oriented dipoles. 
The latter have no effect on the confining properties so the two gases
are identical for confinement at sufficiently large distances. 
So if we are to compare two Gribov copies with respect to those
properties that are responsible for confinement, 
we should really compare the pattern of fluxes in
the two configurations rather than just the numbers and locations of
the monopoles.

The most direct method to do this is to subtract the fluxes
from each other and calculate the Wilson loop with respect to this
{\em difference} of magnetic fluxes:
$$
W^{\diff}({\cal C}) = e^{i(B_1({\cal C})-B_2({\cal C}))}
$$
where $B_1$ and $B_2$ are the magnetic fluxes, through the same
contour $\cal C$, in the two different field configurations.  
So if we want to know whether different Gribov copies of the same
$SU(2)$ gauge field have the same confining fluctuations, we can take
two copies from each $SU(2)$ field, calculate the Wilson loop from the
difference of the fluxes, and average this over an ensemble of $SU(2)$
fields.  
If the resulting expectation value decreases less rapidly than
exponentially with the area, this tells us that the fluctuations in
the differences of the fluxes correspond to a theory with zero string
tension and that Gribov copies have identical confining
fluctuations. 
A non-zero area term provides us with a non-zero `difference' string
tension, $\sigma^{\diff}$.  
By comparing this with $2\sigma$, which is what we would find if the
fluctuations in the two Gribov copies were completely uncorrelated,
we can say whether the effect is `small' or `large'.

In this section we will describe calculations in which
we compare the confining fluctuations of Gribov copies
in the above way. 
We can do so separately for the $U(1)$ fluxes and for the monopole
fluxes. 

If the $U(1)$ link angles of the two Gribov copies are $\setl
\theta^1_l \setr$ and $\setl \theta^2_l \setr$ then we form the
difference links $\theta^{\diff}_l \equiv \theta^1_l - \theta^2_l$ and
we use these to calculate Wilson loops.  
Averaging over an ensemble of $SU(2)$ fields we can obtain the
corresponding potential, $V^{\diff}(r)$ (or Creutz ratios), and 
the string tension, $\sigma^{\diff}$.  

For monopoles we can form a difference gas of monopoles defined by
$m^{\diff}(n)=m_1(n)-m_2(n)$ where $n$ labels the cube and $m$ is the
magnetic charge in that cube.
(This is for D=3; in D=4 one subtracts the currents on the dual
lattice.)   
An alternative procedure is to
extract the monopoles directly from the difference links, 
${\theta^{\diff}_l}$, in the usual way. We call the latter
monopole gas $m^{\diff}_o$. These two difference gases should certainly
have the same long-distance properties, although the sub-leading
contributions might be quite different (as indeed will turn out
to be the case).
We calculate the dual potential for the difference gas and hence the
fluxes through Wilson loops. 
From this we can again extract potentials (or Creutz
ratios) and string tensions. 
In the monopole picture of confinement one would, of course, expect 
to find that the $U(1)$ and monopole string tensions were equal.

We note that this technique of directly comparing fields has a
practical advantage over the calculations described in the previous
section, in that the correlations between the fluctuations in the
different copies are properly taken care of so that the statistical
errors on comparative quantities will be both more accurate and
(probably) smaller.

We begin with our $3$ dimensional calculations and, as in the previous
section, we shall only compare the $U(1)$ fields.  
We have performed calculations with $N_{GT}=2$ on a $16^3$ lattice at
$\beta_3=5$ and on a $24^3$ lattice at $\beta_3=9$. 
In addition we also have the calculations described previously, which
were on a $12^3$ lattice at $\beta_3=5$ with $N_{GT}=30$.  
In physical units the $24^3$ lattice at $\beta_3=9$ is about the same
size as the $12^3$ lattice at $\beta_3=5$. 
Thus we have some check on both the scaling and volume dependence of
our results.

For each $SU(2)$ field we take the pair of Gribov copies that we have
generated and calculate the difference angles, $\setl \theta^{\diff}_l
\setr$, as defined above. 
From these we calculate Wilson loops and from the averages of the
Wilson loops we obtain Creutz ratios. 
From the latter we extract effective (difference) string tensions,
$\sigma^{\diff}_{\eff}(r)$, as in the previous section.  
These are listed in Table~\ref{16_and_24_3} for the $16^3$ and $24^3$
lattices.
For comparison the string tensions one obtains from randomly chosen
gauge copies (the ensemble $\setl R_{av} \setr$ of the previous
section) are $\sigma = 0.087(3)$ and $\sigma = 0.0240(5)$
respectively.  
We see from Table~\ref{16_and_24_3} that $\sigma^{\diff}_{\eff}(r)$
decreases as we go to larger $r$.
At some point the signal gets lost in the noise so it not possible for
us to say whether it goes to zero or not. 
What we can safely do is to put an upper bound
$$
\sigma^{\diff} \le {1 \over 5} \sigma
$$
on the string tension of the `difference' fields.

In Table~\ref{ngt_30_3} we return to our $12^3$ lattice at
$\beta_3=5$.
We show in the first column the values of $\sigma^{\diff}_{\eff}(r)$ 
obtained from randomly chosen pairs of Gribov copies.
In the second column we show the corresponding values when the
difference field is formed by subtracting the $U(1)$ field with the
largest value of $R$ amongst our 30 gauge copies, from the field with
the smallest value of $R$. 
We see a qualitatively similar pattern to that in
Table~\ref{16_and_24_3}, albeit with larger statistical errors.
(Calculations with 30 gauge fixings per $SU(2)$ field are
computationally very expensive and this limits our statistics.)  
As one would expect, at each value of $r$ the difference between
extreme Gribov copies is larger than that between randomly chosen
ones. 
The difference, however, decreases with $r$ and is consistent with going
to zero; in any case we can certainly conclude that
$\sigma^{\diff} \le {1 \over 3} \sigma$.

For purposes of comparison, it is interesting to see what
$\sigma^{\diff}_{\eff}(r)$ looks like if we construct our $U(1)$
difference angles from two fields that do not come from the same
$SU(2)$ field, but instead come from two independent $SU(2)$
fields. 
In practice our `independent' fields are ones which are separated by
50 Monte Carlo sweeps. 
We perform this calculation on $24^3$ fields at $\beta_3=9$, where our
calculations are the most accurate.  
In Figure~\ref{diff_indep2_3} we display
the values of $\sigma^{\diff}_{\eff}(r)$ as calculated in this way.
We also show the string tension extracted from the difference of two
Gribov copies that come from the same $SU(2)$ field (i.e. the values
listed in the appropriate column of Table~\ref{16_and_24_3}) and, in
addition, the effective string tension extracted simply from a single
randomly chosen gauge copy of each $SU(2)$ field. 

The first thing we note is that in this last case the effective string
tension rapidly becomes independent of $r$, making the extraction of
the desired $r \to \infty$ limit straightforward. 
The second thing we note is that this effective string tension is
exactly half the difference string tension calculated from pairs of
independent $SU(2)$ fields. 
This is what one expects for independent fields, where the
fluxes are independent and $\langle W^{\diff} \rangle$ factorises.
Finally we observe that the values of the effective string tension as
calculated from the difference of two Gribov copies from the same
field, are very much smaller and are still decreasing at values of $r$
where the other two effective string tensions have already become
independent of $r$.  
This is our best $D=3$ calculation, with respect to both statistical 
and systematic errors. 
It is consistent with the {\em pattern} of the long-distance fluctuations
which produce confinement, being exactly the same on different Gribov
copies.

In 4 dimensions we are able to obtain useful calculations
of Wilson loops not only from the differences of $U(1)$ 
fields, but also from differences of monopole 
current distributions, as described above. 
In neither case are we able to extract
potentials and string tensions unambiguously. For
the $U(1)$ fields the reason is the same as before.
For the monopoles the reason is that 
the Wilson loops calculated with difference gases 
possess much larger sub-leading contributions than 
with the individual monopole gases.
We shall therefore use Creutz ratios to extract
values of $\sigma^{\diff}_{\eff}$ in all cases.

Our calculations have been performed 
on $12^4$ lattices at $\beta_4 =2.3$ and 2.4.
In Fig~\ref{b_du1} we plot the values of $\sigma^{\diff}_{\eff}$ 
that we have obtained from the $U(1)$ difference 
fields that have been extracted from pairs of 
randomly chosen Gribov copies from each $SU(2)$
field configuration. They are shown as a function of
the area of the (largest) Wilson loop used in the
evaluation of the Creutz ratio. We have included
not only square (largest) loops, as in the previous
section, but also some rectangular loops. (Although
we have limited ourselves to using ones which are
nearly square.) 
We see from Fig~\ref{b_du1} that $\sigma^{\diff}_{\eff}$ drops
smoothly towards zero with increasing area. Whether
it actually asymptotes to zero or not is something 
that we cannot say, because of the increasing
errors at larger areas. Nonetheless what we can safely 
infer is that the asymptotic $U(1)$ difference string
tension is bounded by $\sigma^{\diff} \le {2\over 5}\sigma$. 
 
We now extract the monopole gases from the $U(1)$ difference
fields of the previous paragraph, and calculate
the corresponding Creutz ratios. The resulting values of 
$\sigma^{\diff}_{\eff}$ are plotted in Fig~\ref{b_dgtd}. We also show on 
these plots the $\sigma_{\eff}$ one obtains from the
original monopole gases at the corresponding values of
$\beta_4$. We observe that $\sigma^{\diff}_{\eff}$ drops
monotonically towards zero. This is especially striking
when compared to the behaviour of $\sigma_{\eff}$. 
From these figures we obtain a typical bound
$\sigma^{\diff} \le {2\over 5}\sigma$. 

As mentioned earlier, an alternative to the above is to 
subtract the the two monopole gases so as to produce a difference 
monopole gas. From this we can calculate Creutz ratios and 
these are displayed in Fig~\ref{b_ddgt}. Comparing to Fig~\ref{b_dgtd}
we see that these effective string tensions decrease
much more slowly as a function of loop area. It
appears that this method of comparing monopole 
properties greatly enhances the sub-leading shorter
distance fluctuations (for reasons that we have not
completely understood). The bound one obtains from
Fig~\ref{b_ddgt} on $\sigma^{\diff}$ is clearly weaker than
our previous bounds and this method is clearly much less efficient.

We have found that in both 3 and 4 dimensions the
long-distance fluctuations that drive linear confinement
are remarkably similar in pairs of randomly chosen
Gribov copies. 

This result naturally provokes the question: perhaps
the difference between the monopole gases on different
Gribov copies is manifestly trivial? For example it 
might be due to monopole-antimonopole pairs one or two
lattice spacings apart. 

One way of probing this question is to calculate the lengths 
of the monopole loops and to examine the number of loops of each
length (the loop `spectrum'). If the monopole difference gas were
trivial then we would expect a spectrum greatly more peaked
at small loop lengths than that of either of the configurations
contributing to the difference gas. In fact this is not what we find.
Instead we obtain a spectrum which extends to large loops.
Indeed, apart from the overall normalisation, we find
that this spectrum is indistinguishable from that
which we obtain in the usual monopole gas. An analogous
conclusion is reached in 3 dimensions. Thus the difference
is certainly not trivial. A natural possibility is that it might 
consist of dipoles whose extent is limited by some physical
length scale, such as the extent of the 't Hooft-Polyakov 
cores in the simple `model' we introduced earlier. 
With a separation of several lattice spacings, any
constraint on the loop lengths would be much less severe.
This is a much harder possibility to test and we shall
not attempt to do so here.

\section	{Conclusions.}
\label		{conclusion}

The Maximally Abelian gauge has provided a promising framework within
which to explore the monopole approach to confinement.  
There are, however, Gribov copies in this gauge and quantities such as
the monopole density vary quite strongly between the different copies
of the same non-Abelian gauge field. 
This naturally raises the question of what significance 
should be attached to the
observation that, if one performs calculations on the $U(1)$ fields
that are extracted from randomly chosen Gribov copies, then the
Abelian and non-Abelian string tensions are approximately equal.

In this paper we have confirmed that Gribov copies differ
in various respects, such as the monopole density. 
We have however argued that this need not be an important difference,
and we illustrated this within a dynamical `picture' where confinement
arises from the condensation of 't Hooft-Polyakov monopoles following
the formation of composite scalars and a dynamical symmetry breaking
in the non-Abelian fields.

We then attempted to establish in a direct fashion whether or not 
the confining  properties of different Gribov copies are the same.
We first calculated the $U(1)$ and monopole 
(effective) string tensions on Gribov
copies that had been selected into separate ensembles on the basis of
the value of the quantity $R$ which is maximised when one fixes to the
MA gauge. 
What we found was that the (effective) string tension does indeed 
vary, and that it decreases as we increase the value of $R$. 
This variation is, however, small compared to the typical
value of the string tension. For example, we find that the string 
tension calculated on Gribov copies with the largest values of $R$
is about $10\%$ less than that obtained with randomly 
chosen Gribov copies.   

Comparing string tensions on such selected ensembles of Gribov copies
provides a rather crude probe of the differences between such
copies. 
What one really wants to do is to compare, between different Gribov
copies of the same non-Abelian field, those $U(1)$ field fluctuations
that drive confinement. 
We suggested that this could be achieved most simply by subtracting the
Abelian gauge potentials of two copies and calculating Wilson loops on
this new field which is the difference of the old ones. 
If the pattern of long-distance confining fluctuations is identical on
different Gribov copies, then the resulting `difference' string
tension, $\sigma^{\diff}$, should be zero. 
If the fluctuations are completely uncorrelated then we should find
$\sigma^{\diff} = 2 \sigma$, where $\sigma$ is the string tension one
obtains from Wilson loop calculations on randomly chosen Gribov
copies.  
What we actually found, in both 3 and 4 dimensions, is that
$\sigma^{\diff} \le \frac{1}{6} 2\sigma$ and, indeed, is compatible 
with zero. 
That is to say, as far as the confining fluctuations are
concerned, different Gribov copies of the same non-Abelian
field are very strongly correlated.

The fact that there does appear to be some variation
of $\sigma$ with $R$, has further implications. Consider
the calculation of the string tension in the ensemble of
randomly chosen Gribov copies. We have assumed that there is
a particular value of $\sigma$ associated to this ensemble.
However, by definition, this ensemble contains Gribov copies
with all possible values of $R$. Thus one would na\"{\i}vely expect 
that Wilson loops of large enough area $A$ 
will be given by
$$
\langle W(A) \rangle \sim \int \limits_{\sigma_{-}}^{\sigma_{+}}
d\sigma \rho(\sigma) e^{-\sigma A} 
$$
where $\rho(\sigma)$ is some suitable density function which takes
into account the fact that there is a variation of $\sigma$ with $R$
between limits $\sigma_{-}$ and $\sigma_{+}$. Since we observe that
the value of $\sigma$ appears to decrease monotonically with
increasing $R$, we may assume that $\sigma_{-}$ is the value of
$\sigma$ for the Gribov copies with the largest values of $R$. Now we
observe that
$$
\lim_{A\to\infty} \langle W(A) \rangle \sim 
e^{-\sigma_{-} A} 
$$
However if the variation of $\sigma$ with $R$ is as small as we
find it to be, one can easily estimate that this
`true' asymptotic behaviour would only become visible
for Wilson loops that are far larger than those we
have been able to consider. Thus it is no surprise
that we have not observed this effect. Nonetheless
the implication is that
the `true' asymptotic value of $\sigma$, as obtained
with randomly chosen Gribov copies, is in fact the value
of $\sigma$ that one would obtain if one were to
use the Gribov copies with the largest value of $R$.
Clearly the most efficient way to calculate this
string tension is to choose such extreme Gribov copies from 
the start. This provides a concrete argument for 
focusing on these extreme Gribov copies.

Of course one must be careful with the `na\"{\i}ve' argument
of the previous paragraph. The basic assumption was that
there is a different $\sigma$ for different $R$, so that
the string tension we obtain with randomly chosen Gribov
copies is a sub-leading phenomenon that will 
be transformed, for sufficiently large areas, into
$\sigma_{-}$. It might be that the situation is the
reverse, i.e. there is a definite string tension, $\bar \sigma$,
associated with randomly chosen Gribov copies and the
$R$-dependent values of $\sigma$ that we observe are in
fact a sub-leading phenomenon and will 
be transformed for sufficiently large areas into
$\bar \sigma$. Even if our first assumption is correct,
there are some obvious questions that need to be answered
such as what is the large volume dependence of $\rho(\sigma)$, 
how does the onset of the area law vary with $R$ etc., 
before we can take the conclusion of the previous
paragraph too seriously.

We conclude that even though we see a modest variation of
the string tension with the particular ensemble of Gribov copies
used, the confining Abelian fluctuations of different
Gribov copies are in fact remarkably correlated. This
is so in the zero temperature confining phase for both 3 and 4
dimensions.  
It suggests that while Gribov copies do indeed pose a 
real problem in principle - for the study of confinement 
in the Maximally Abelian gauge - the 
practical ambiguities are really quite limited and do not
seriously undermine the evidence that has made this
gauge so interesting. 

\vskip 0.50in

{\bf Note Added}

When this work was completed we received a paper 
\cite{Balinew} in which
techniques are developed to calculate string tensions on
the Gribov copies corresponding to the absolute maximum
of $R$. These authors also observe that $\sigma$
decreases as $R$ increases.
\section*	{Acknowledgments}

We thank Gunnar Bali, Valja Mitrjushkin, Klaus Schilling, Zsolt 
Schram and John Stack for interesting discussions during the course
of this work.
One of us (AGH) acknowledges the support of a PPARC studentship.
The calculations were supported by PPARC Grants GR/J21408 and
GR/K55752.

\newpage

\begin 	{table}[p]
\begin	{center}
\begin	{tabular}
{||r@{.}l|| r@{.}l@{ (}r@{) }| r@{.}l@{ (}r@{) }| r@{.}l@{ (}r@{) }||}
\hline \hline
\multicolumn{2}{||c||}{$\beta_3$} & \multicolumn{3}{|c|}{$f$} & 
\multicolumn{3}{|c|}{$p$} & \multicolumn{3}{|c||}{$\ng$} \\
\hline
 4&0 & 	0&128 & 25 & 	0&35 & 11 & 	 28&0  &  17 \\
 5&0 & 	0&363 & 37 & 	0&80 & 10 & 	 14&8  &  15 \\
 6&0 & 	0&684 & 59 & 	0&95 &  5 & 	  5&7  &   9 \\
 7&0 & 	0&799 & 45 & 	1&00 &  5 & 	  4&1  &   8 \\
 8&0 & 	0&938 & 31 & 	1&00 &  5 & 	  1&55 &  20 \\
 9&0 & 	0&972 & 10 & 	1&00 &  5 & 	  1&60 &  20 \\
10&0 & 	0&983 &  8 & 	1&00 &  5 & 	  1&35 &  17 \\
11&0 & 	0&989 &  5 & 	1&00 &  5 & 	  1&25 &  10 \\
12&0 & 	0&986 &  5 & 	1&00 &  5 & 	  1&35 &  11 \\
\hline \hline
\end	{tabular}
\end	{center}
\caption{Some properties of Gribov copies on an $8^3$ lattice in $D=3$.}
\label	{f_p_ng_3}
\end 	{table}

\begin 	{table}[p]
\begin	{center}
\begin	{tabular}
{||r@{.}l|| r@{.}l@{ (}r@{) }| r@{.}l@{ (}r@{) }| r@{.}l@{ (}r@{) }||}
\hline \hline
\multicolumn{2}{||c||}{$\beta_4$} & \multicolumn{3}{|c|}{$f$} & 
\multicolumn{3}{|c|}{$p$} & \multicolumn{3}{|c||}{$\ng$} \\
\hline
2&3 &	0&01 &  2 &	0&25 &  5 &	98&6  &  11 \\
2&4 & 	0&07 & 17 &	0&35 &  5 &	77&8  & 252 \\
2&5 & 	0&43 & 33 &	0&70 &  5 &	22&1  & 266 \\
2&6 & 	0&68 & 38 &	0&75 &  5 &	 8&6  & 139 \\
2&7 & 	0&91 & 16 &	1&00 &  5 &	 2&1  &  19 \\
\hline \hline
\end	{tabular}
\end	{center}
\caption{Some properties of Gribov copies on an $8^4$ lattice in $D=4$.}
\label	{f_p_ng_4}
\end 	{table}

\begin 	{table}[p]
\begin	{center}
\begin	{tabular}
{||c|| r@{.}l@{ (}r@{) }| r@{.}l@{ (}r@{) }| r@{.}l@{ (}r@{) }||}
\cline{2-10}
\multicolumn{1}{c|}{} & \multicolumn{3}{|c|}{$R_{min}$} & 
\multicolumn{3}{|c|}{$R_{av}$} & \multicolumn{3}{|c||}{$R_{max}$} \\
\hline
$\langle R \rangle$ &
0&8477 & 3 &	0&8533 & 4 &	0&8566 & 3 \\
\hline
$\langle \cos \theta_p \rangle$ & 	
0&8856 & 5 &	0&8884 & 4 &	0&8901 & 5 \\
\hline
$ \langle n_{\hbox{\tiny M}} \rangle $ &
5&875 & 83 &	4&752 & 44 &	4&018 & 72 \\
\hline \hline
\end	{tabular}
\end	{center}
\caption{Average $R$, $U(1)$ plaquette and monopole number in $D=3$ on a
$12^3$ lattice at $\beta_3 = 5$, using $N_{GT}=30$.}
\label	{cos_nm_3}
\end 	{table}

\begin 	{table}[p]
\begin	{center}
\begin	{tabular}
{||c|| r@{.}l@{ (}r@{) }| r@{.}l@{ (}r@{) }| r@{.}l@{ (}r@{) }||}
\hline \hline
& \multicolumn{9}{|c||}{$\sigma_{\eff} (r)$} \\
\cline{2-10} 
$r$ & \multicolumn{3}{|c|}{$R_{min}$} & \multicolumn{3}{|c|}{$R_{av}$} & 
\multicolumn{3}{|c||}{$R_{max}$} \\
\hline
2 & 	0&1078 & 15 & 	0&0999 &  8 & 	0&0957 & 13 \\
3 &	0&1028 & 30 &	0&0916 & 18 &	0&0841 & 23 \\	
4 &	0&107  &  7 &	0&089  &  3 &	0&079  &  5 \\
5 &	0&092  & 16 &	0&083  &  7 &	0&079  & 11 \\
\hline \hline
\end	{tabular}
\end	{center}
\caption{Effective $U(1)$ string tensions in $D=3$ on a $12^3$
lattice at $\beta_3 = 5$, using $N_{GT}=30$.}
\label	{sig_grib_3}
\end 	{table}

\begin 	{table}[p]
\begin	{center}
\begin	{tabular}
{||c| r@{.}l@{ (}r@{) }| r@{.}l@{ (}r@{) }| r@{.}l@{ (}r@{) }||}
\cline{2-10}
\multicolumn{1}{c|}{} & \multicolumn{3}{|c|}{$R_{min}$} & 
\multicolumn{3}{|c|}{$R_{av}$} & \multicolumn{3}{|c||}{$R_{max}$} \\
\hline
$\langle R \rangle$ &
0&7297 & 1 &	0&7308 & 1 &	0&7320 & 1 \\
\hline
$\langle \cos \theta_p \rangle$ & 	
0&7066 & 5 &	0&7094 & 7 &	0&7132 & 5 \\
\hline
$ \langle P_{\hbox{\tiny M}} \rangle $ &
0&0541 & 4 &	0&0514 & 6 &	0&0489 & 4 \\
\hline \hline
\end	{tabular}
\end	{center}
\caption{Average $R$, $U(1)$ plaquette and summed lengths of monopole loops
(normalised by the number of lattice links) in $D=4$ on a $12^4$
lattice at $\beta_4 = 2.4$, using $N_{GT}=5$.}
\label	{cos_perim_4}
\end 	{table}

\begin 	{table}[p]
\begin	{center}
\begin	{tabular}
{||c|| r@{.}l@{ (}r@{) }| r@{.}l@{ (}r@{) }| r@{.}l@{ (}r@{) }||}
\hline \hline
& \multicolumn{9}{|c||}{$\sigma_{\eff} (r)$} \\
\cline{2-10} 
$r$ & \multicolumn{3}{|c|}{$R_{min}$} & \multicolumn{3}{|c|}{$R_{av}$} & 
\multicolumn{3}{|c||}{$R_{max}$} \\
\hline
2 &	0&230 &	 3  &	0&224 &	 3  &	0&223 &  3 \\
3 &	0&176 &	 9  &	0&190 &	10  &	0&162 & 10 \\
4 &	0&128 &	39  &	0&113 &	47  &	0&161 & 41 \\
\hline \hline
\end	{tabular}

\vspace{3ex}
\begin	{tabular}
{||c|| r@{.}l@{ (}r@{) }| r@{.}l@{ (}r@{) }| r@{.}l@{ (}r@{) }||}
\hline \hline
& \multicolumn{9}{|c||}{$\sigma_{\eff} (r)$} \\
\cline{2-10} 
$r$ & \multicolumn{3}{|c|}{$R_{min}$} & \multicolumn{3}{|c|}{$R_{av}$} & 
\multicolumn{3}{|c||}{$R_{max}$} \\
\hline
2 &	0&159 &	 3  &	0&160 &	 2  &	0&151 &  3 \\
3 &	0&116 &	 6  &	0&112 &	 4  &	0&097 &  7 \\
4 &	0&105 &	20  &	0&085 &	11  &	0&100 & 13 \\
5 &	0&166 &	73  &	0&069 &	43  &	0&120 & 48 \\
\hline \hline
\end	{tabular}
\end	{center}
\caption{Effective $U(1)$ string tensions in $D=4$ on a $12^4$
lattice, using $N_{GT} = 10$  at $\beta_4 = 2.3$ and 
$N_{GT} = 5$  at $\beta_4 = 2.4$ respectively.}
\label	{sig_u1_4}
\end 	{table}

\begin 	{table}[p]
\begin	{center}
\begin	{tabular}
{||c| r@{.}l@{ (}r@{) }| r@{.}l@{ (}r@{) }| r@{.}l@{ (}r@{) }||}
\hline \hline
& \multicolumn{9}{|c||}{$\sigma$} \\
\cline{2-10}
$\beta_4$ &
\multicolumn{3}{|c|}{$R_{min}$} & \multicolumn{3}{|c|}{$R_{av}$} & 
\multicolumn{3}{|c||}{$R_{max}$} \\
\hline
2.3 & 	0&136  &  2 &	0&127  &  2 &	0&121  &  2 \\
\hline
2.4 &	0&068  &  2 &	0&064  &  2 &	0&058  &  2 \\
\hline \hline
\end	{tabular}
\end	{center}
\caption{Monopole string tensions in $D=4$ on a $12^4$
lattice, using $N_{GT} = 10$  at $\beta_4 = 2.3$ and 
$N_{GT} = 5$  at $\beta_4 = 2.4$.}
\label	{sig_mon_4}
\end 	{table}

\begin 	{table}[p]
\begin	{center}
\begin	{tabular}
{||c|| r@{.}l@{ (}r@{) }| r@{.}l@{ (}r@{) }||}
\hline \hline
& \multicolumn{6}{|c||}{$\sigma^{\diff}_{\eff} (r)$} \\
\cline{2-7} 
$r$ & \multicolumn{3}{|c|}{$ 16^3$; $\beta_3 = 5$} & 
\multicolumn{3}{|c||}{$ 24^3$; $\beta_3 = 9$} \\
\hline
 2 &	 0&1106 &  12 &		 	 0&0549 &  4 \\
 3 &	 0&0387 &  13 &		 	 0&0320 &  5 \\
 4 &	 0&0158 &  23 &		 	 0&0195 &  6 \\
 5 &	 0&0127 &  51 &		 	 0&0103 & 10 \\
 6 &	 0&0122 &  95 &		 	 0&0085 & 14 \\
 7 &	-0&0150 & 110 &		 	 0&0055 & 20 \\
 8 &	\multicolumn{3}{|c|}{} &	 0&0018 & 27 \\
 9 &	\multicolumn{3}{|c|}{}	& 	-0&0013 & 31 \\
10 &	\multicolumn{3}{|c|}{}	&  	 0&0060 & 62 \\
\hline \hline
\end	{tabular}
\end	{center}
\caption{Effective $U(1)$ string tensions of difference gases between
random Gribov copies in $D=3$ on lattices of comparable physical
sizes.}
\label	{16_and_24_3}
\end 	{table}

\begin 	{table}[p]
\begin	{center}
\begin	{tabular}
{||c|| r@{.}l@{ (}r@{) }| r@{.}l@{ (}r@{) }||}
\hline \hline
& \multicolumn{6}{|c||}{$\sigma^{\diff}_{\eff} (r)$} \\
\cline{2-7} 
$r$ & \multicolumn{3}{|c|}{$ N_{\hbox{\tiny GT}} = 2 $} & 
\multicolumn{3}{|c||}{$ N_{\hbox{\tiny GT}} = 30 $} \\
\hline
2 &	 0&1116 & 33 &		0&1209 &  26 \\
3 &	 0&0351 & 40 &		0&0558 &  50 \\
4 &	 0&0111 & 90 &		0&0243 & 100 \\
5 &	-0&013  & 15 &		0&0040 & 200 \\
\hline \hline
\end	{tabular}
\end	{center}
\caption{Effective $U(1)$ string tensions of difference gases between
Gribov copies that are extreme in $R$ using $N_{GT}$ gauge transformations 
in $D=3$ on a $ 12^3$ lattice at $\beta_3 = 5$.}
\label	{ngt_30_3}
\end 	{table}

\begin 	{figure}[p]
\begin	{center}
\leavevmode
\input{b9l24_3}
\end	{center}
\caption{Comparison of $U(1)$ effective string tensions in $D=3$ on a $24^3$
lattice at $\beta_3 = 9$. 
Key: $+$~full $U(1)$ fields ($R_{av}$), $\times$~Difference fields from
independent configurations,  $\star$~Difference fields from random
Gribov copies.}
\label	{diff_indep2_3}
\end 	{figure}
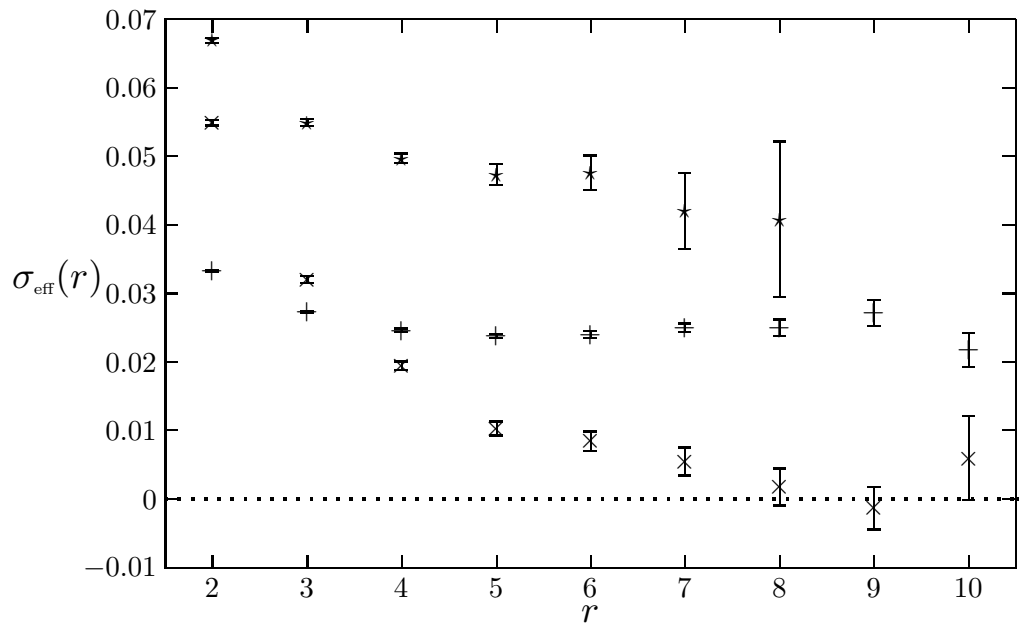

\begin{figure}[p]
\begin{center}
\leavevmode
\input{b23_du1}

\vspace{10ex}
\input{b24_du1}
\end{center}
\caption{Effective string tensions for the $U(1)$ difference fields from
extremal Gribov copies at
$\beta_4 = 2.3, 2.4$ respectively.}
\label{b_du1}
\end{figure}

\begin{figure}[p]
\begin{center}
\leavevmode
\input{b23_dgtd}

\vspace{10ex}
\input{b24_dgtd}
\end{center}
\caption{Effective string tensions for; $+$~monopoles,
$\times$~monopoles identified from the $U(1)$ difference fields from
extremal Gribov copies, at $\beta_4 = 2.3, 2.4$ respectively}
\label{b_dgtd}
\end{figure}

\begin{figure}[p]
\begin{center}
\leavevmode
\input{b23_ddgt}

\vspace{10ex}
\input{b24_ddgt}
\end{center}
\caption{Effective string tensions for the monopole difference gases
from extremal Gribov copies at
$\beta_4 = 2.3, 2.4$ respectively.}
\label{b_ddgt}
\end{figure}

\end{document}

%% file: b9l24_3.tex
\setlength{\unitlength}{0.240900pt}
\ifx\plotpoint\undefined\newsavebox{\plotpoint}\fi
\sbox{\plotpoint}{\rule[-0.200pt]{0.400pt}{0.400pt}}%
\begin{picture}(1500,900)(0,0)
\font\gnuplot=cmr10 at 12pt
\gnuplot
\sbox{\plotpoint}{\rule[-0.200pt]{0.400pt}{0.400pt}}%
\put(120.0,31.0){\rule[-0.200pt]{4.818pt}{0.400pt}}
\put(108,31){\makebox(0,0)[r]{$-0.01$}}
\put(1436.0,31.0){\rule[-0.200pt]{4.818pt}{0.400pt}}
\put(120.0,139.0){\rule[-0.200pt]{4.818pt}{0.400pt}}
\put(108,139){\makebox(0,0)[r]{$0$}}
\put(1436.0,139.0){\rule[-0.200pt]{4.818pt}{0.400pt}}
\put(120.0,247.0){\rule[-0.200pt]{4.818pt}{0.400pt}}
\put(108,247){\makebox(0,0)[r]{$0.01$}}
\put(1436.0,247.0){\rule[-0.200pt]{4.818pt}{0.400pt}}
\put(120.0,354.0){\rule[-0.200pt]{4.818pt}{0.400pt}}
\put(108,354){\makebox(0,0)[r]{$0.02$}}
\put(1436.0,354.0){\rule[-0.200pt]{4.818pt}{0.400pt}}
\put(120.0,462.0){\rule[-0.200pt]{4.818pt}{0.400pt}}
\put(108,462){\makebox(0,0)[r]{$0.03$}}
\put(1436.0,462.0){\rule[-0.200pt]{4.818pt}{0.400pt}}
\put(120.0,570.0){\rule[-0.200pt]{4.818pt}{0.400pt}}
\put(108,570){\makebox(0,0)[r]{$0.04$}}
\put(1436.0,570.0){\rule[-0.200pt]{4.818pt}{0.400pt}}
\put(120.0,678.0){\rule[-0.200pt]{4.818pt}{0.400pt}}
\put(108,678){\makebox(0,0)[r]{$0.05$}}
\put(1436.0,678.0){\rule[-0.200pt]{4.818pt}{0.400pt}}
\put(120.0,785.0){\rule[-0.200pt]{4.818pt}{0.400pt}}
\put(108,785){\makebox(0,0)[r]{$0.06$}}
\put(1436.0,785.0){\rule[-0.200pt]{4.818pt}{0.400pt}}
\put(120.0,893.0){\rule[-0.200pt]{4.818pt}{0.400pt}}
\put(108,893){\makebox(0,0)[r]{$0.07$}}
\put(1436.0,893.0){\rule[-0.200pt]{4.818pt}{0.400pt}}
\put(194.0,31.0){\rule[-0.200pt]{0.400pt}{4.818pt}}
\put(194,19){\makebox(0,0){\shortstack{\\ \\ \\ $2$}}}
\put(194.0,873.0){\rule[-0.200pt]{0.400pt}{4.818pt}}
\put(343.0,31.0){\rule[-0.200pt]{0.400pt}{4.818pt}}
\put(343,19){\makebox(0,0){\shortstack{\\ \\ \\ $3$}}}
\put(343.0,873.0){\rule[-0.200pt]{0.400pt}{4.818pt}}
\put(491.0,31.0){\rule[-0.200pt]{0.400pt}{4.818pt}}
\put(491,19){\makebox(0,0){\shortstack{\\ \\ \\ $4$}}}
\put(491.0,873.0){\rule[-0.200pt]{0.400pt}{4.818pt}}
\put(640.0,31.0){\rule[-0.200pt]{0.400pt}{4.818pt}}
\put(640,19){\makebox(0,0){\shortstack{\\ \\ \\ $5$}}}
\put(640.0,873.0){\rule[-0.200pt]{0.400pt}{4.818pt}}
\put(788.0,31.0){\rule[-0.200pt]{0.400pt}{4.818pt}}
\put(788,19){\makebox(0,0){\shortstack{\\ \\ \\ $6$}}}
\put(788.0,873.0){\rule[-0.200pt]{0.400pt}{4.818pt}}
\put(936.0,31.0){\rule[-0.200pt]{0.400pt}{4.818pt}}
\put(936,19){\makebox(0,0){\shortstack{\\ \\ \\ $7$}}}
\put(936.0,873.0){\rule[-0.200pt]{0.400pt}{4.818pt}}
\put(1085.0,31.0){\rule[-0.200pt]{0.400pt}{4.818pt}}
\put(1085,19){\makebox(0,0){\shortstack{\\ \\ \\ $8$}}}
\put(1085.0,873.0){\rule[-0.200pt]{0.400pt}{4.818pt}}
\put(1233.0,31.0){\rule[-0.200pt]{0.400pt}{4.818pt}}
\put(1233,19){\makebox(0,0){\shortstack{\\ \\ \\ $9$}}}
\put(1233.0,873.0){\rule[-0.200pt]{0.400pt}{4.818pt}}
\put(1382.0,31.0){\rule[-0.200pt]{0.400pt}{4.818pt}}
\put(1382,19){\makebox(0,0){\shortstack{\\ \\ \\ $10$}}}
\put(1382.0,873.0){\rule[-0.200pt]{0.400pt}{4.818pt}}
\put(120.0,31.0){\rule[-0.200pt]{321.842pt}{0.400pt}}
\put(1456.0,31.0){\rule[-0.200pt]{0.400pt}{207.656pt}}
\put(120.0,893.0){\rule[-0.200pt]{321.842pt}{0.400pt}}
\put(-48,486){\makebox(0,0){{\Large $\sigma_{\eff} (r)$}}}
\put(788,-41){\makebox(0,0){{\Large $r$}}}
\put(120.0,31.0){\rule[-0.200pt]{0.400pt}{207.656pt}}
\put(194,498){\makebox(0,0){$+$}}
\put(343,433){\makebox(0,0){$+$}}
\put(491,404){\makebox(0,0){$+$}}
\put(640,395){\makebox(0,0){$+$}}
\put(788,397){\makebox(0,0){$+$}}
\put(936,408){\makebox(0,0){$+$}}
\put(1085,408){\makebox(0,0){$+$}}
\put(1233,432){\makebox(0,0){$+$}}
\put(1382,373){\makebox(0,0){$+$}}
\put(194.0,496.0){\rule[-0.200pt]{0.400pt}{0.723pt}}
\put(184.0,496.0){\rule[-0.200pt]{4.818pt}{0.400pt}}
\put(184.0,499.0){\rule[-0.200pt]{4.818pt}{0.400pt}}
\put(343.0,431.0){\rule[-0.200pt]{0.400pt}{0.964pt}}
\put(333.0,431.0){\rule[-0.200pt]{4.818pt}{0.400pt}}
\put(333.0,435.0){\rule[-0.200pt]{4.818pt}{0.400pt}}
\put(491.0,401.0){\rule[-0.200pt]{0.400pt}{1.445pt}}
\put(481.0,401.0){\rule[-0.200pt]{4.818pt}{0.400pt}}
\put(481.0,407.0){\rule[-0.200pt]{4.818pt}{0.400pt}}
\put(640.0,392.0){\rule[-0.200pt]{0.400pt}{1.445pt}}
\put(630.0,392.0){\rule[-0.200pt]{4.818pt}{0.400pt}}
\put(630.0,398.0){\rule[-0.200pt]{4.818pt}{0.400pt}}
\put(788.0,392.0){\rule[-0.200pt]{0.400pt}{2.650pt}}
\put(778.0,392.0){\rule[-0.200pt]{4.818pt}{0.400pt}}
\put(778.0,403.0){\rule[-0.200pt]{4.818pt}{0.400pt}}
\put(936.0,402.0){\rule[-0.200pt]{0.400pt}{3.132pt}}
\put(926.0,402.0){\rule[-0.200pt]{4.818pt}{0.400pt}}
\put(926.0,415.0){\rule[-0.200pt]{4.818pt}{0.400pt}}
\put(1085.0,395.0){\rule[-0.200pt]{0.400pt}{6.263pt}}
\put(1075.0,395.0){\rule[-0.200pt]{4.818pt}{0.400pt}}
\put(1075.0,421.0){\rule[-0.200pt]{4.818pt}{0.400pt}}
\put(1233.0,411.0){\rule[-0.200pt]{0.400pt}{9.877pt}}
\put(1223.0,411.0){\rule[-0.200pt]{4.818pt}{0.400pt}}
\put(1223.0,452.0){\rule[-0.200pt]{4.818pt}{0.400pt}}
\put(1382.0,346.0){\rule[-0.200pt]{0.400pt}{13.009pt}}
\put(1372.0,346.0){\rule[-0.200pt]{4.818pt}{0.400pt}}
\put(1372.0,400.0){\rule[-0.200pt]{4.818pt}{0.400pt}}
\put(194,730){\makebox(0,0){$\times$}}
\put(343,484){\makebox(0,0){$\times$}}
\put(491,349){\makebox(0,0){$\times$}}
\put(640,250){\makebox(0,0){$\times$}}
\put(788,230){\makebox(0,0){$\times$}}
\put(936,198){\makebox(0,0){$\times$}}
\put(1085,158){\makebox(0,0){$\times$}}
\put(1233,125){\makebox(0,0){$\times$}}
\put(1382,203){\makebox(0,0){$\times$}}
\put(194.0,726.0){\rule[-0.200pt]{0.400pt}{2.168pt}}
\put(184.0,726.0){\rule[-0.200pt]{4.818pt}{0.400pt}}
\put(184.0,735.0){\rule[-0.200pt]{4.818pt}{0.400pt}}
\put(343.0,478.0){\rule[-0.200pt]{0.400pt}{2.650pt}}
\put(333.0,478.0){\rule[-0.200pt]{4.818pt}{0.400pt}}
\put(333.0,489.0){\rule[-0.200pt]{4.818pt}{0.400pt}}
\put(491.0,342.0){\rule[-0.200pt]{0.400pt}{3.132pt}}
\put(481.0,342.0){\rule[-0.200pt]{4.818pt}{0.400pt}}
\put(481.0,355.0){\rule[-0.200pt]{4.818pt}{0.400pt}}
\put(640.0,239.0){\rule[-0.200pt]{0.400pt}{5.300pt}}
\put(630.0,239.0){\rule[-0.200pt]{4.818pt}{0.400pt}}
\put(630.0,261.0){\rule[-0.200pt]{4.818pt}{0.400pt}}
\put(788.0,215.0){\rule[-0.200pt]{0.400pt}{7.227pt}}
\put(778.0,215.0){\rule[-0.200pt]{4.818pt}{0.400pt}}
\put(778.0,245.0){\rule[-0.200pt]{4.818pt}{0.400pt}}
\put(936.0,176.0){\rule[-0.200pt]{0.400pt}{10.600pt}}
\put(926.0,176.0){\rule[-0.200pt]{4.818pt}{0.400pt}}
\put(926.0,220.0){\rule[-0.200pt]{4.818pt}{0.400pt}}
\put(1085.0,129.0){\rule[-0.200pt]{0.400pt}{13.972pt}}
\put(1075.0,129.0){\rule[-0.200pt]{4.818pt}{0.400pt}}
\put(1075.0,187.0){\rule[-0.200pt]{4.818pt}{0.400pt}}
\put(1233.0,91.0){\rule[-0.200pt]{0.400pt}{16.140pt}}
\put(1223.0,91.0){\rule[-0.200pt]{4.818pt}{0.400pt}}
\put(1223.0,158.0){\rule[-0.200pt]{4.818pt}{0.400pt}}
\put(1382.0,137.0){\rule[-0.200pt]{0.400pt}{32.040pt}}
\put(1372.0,137.0){\rule[-0.200pt]{4.818pt}{0.400pt}}
\put(1372.0,270.0){\rule[-0.200pt]{4.818pt}{0.400pt}}
\put(194,860){\makebox(0,0){$\star$}}
\put(343,730){\makebox(0,0){$\star$}}
\put(491,674){\makebox(0,0){$\star$}}
\put(640,649){\makebox(0,0){$\star$}}
\put(788,652){\makebox(0,0){$\star$}}
\put(936,591){\makebox(0,0){$\star$}}
\put(1085,578){\makebox(0,0){$\star$}}
\put(194.0,856.0){\rule[-0.200pt]{0.400pt}{1.686pt}}
\put(184.0,856.0){\rule[-0.200pt]{4.818pt}{0.400pt}}
\put(184.0,863.0){\rule[-0.200pt]{4.818pt}{0.400pt}}
\put(343.0,725.0){\rule[-0.200pt]{0.400pt}{2.650pt}}
\put(333.0,725.0){\rule[-0.200pt]{4.818pt}{0.400pt}}
\put(333.0,736.0){\rule[-0.200pt]{4.818pt}{0.400pt}}
\put(491.0,667.0){\rule[-0.200pt]{0.400pt}{3.613pt}}
\put(481.0,667.0){\rule[-0.200pt]{4.818pt}{0.400pt}}
\put(481.0,682.0){\rule[-0.200pt]{4.818pt}{0.400pt}}
\put(640.0,633.0){\rule[-0.200pt]{0.400pt}{7.950pt}}
\put(630.0,633.0){\rule[-0.200pt]{4.818pt}{0.400pt}}
\put(630.0,666.0){\rule[-0.200pt]{4.818pt}{0.400pt}}
\put(788.0,625.0){\rule[-0.200pt]{0.400pt}{13.009pt}}
\put(778.0,625.0){\rule[-0.200pt]{4.818pt}{0.400pt}}
\put(778.0,679.0){\rule[-0.200pt]{4.818pt}{0.400pt}}
\put(936.0,532.0){\rule[-0.200pt]{0.400pt}{28.667pt}}
\put(926.0,532.0){\rule[-0.200pt]{4.818pt}{0.400pt}}
\put(926.0,651.0){\rule[-0.200pt]{4.818pt}{0.400pt}}
\put(1085.0,456.0){\rule[-0.200pt]{0.400pt}{59.020pt}}
\put(1075.0,456.0){\rule[-0.200pt]{4.818pt}{0.400pt}}
\put(1075.0,701.0){\rule[-0.200pt]{4.818pt}{0.400pt}}
\sbox{\plotpoint}{\rule[-0.500pt]{1.000pt}{1.000pt}}%
\put(120,139){\usebox{\plotpoint}}
\put(120.00,139.00){\usebox{\plotpoint}}
\put(140.76,139.00){\usebox{\plotpoint}}
\multiput(147,139)(20.756,0.000){0}{\usebox{\plotpoint}}
\put(161.51,139.00){\usebox{\plotpoint}}
\put(182.27,139.00){\usebox{\plotpoint}}
\multiput(187,139)(20.756,0.000){0}{\usebox{\plotpoint}}
\put(203.02,139.00){\usebox{\plotpoint}}
\put(223.78,139.00){\usebox{\plotpoint}}
\multiput(228,139)(20.756,0.000){0}{\usebox{\plotpoint}}
\put(244.53,139.00){\usebox{\plotpoint}}
\put(265.29,139.00){\usebox{\plotpoint}}
\multiput(268,139)(20.756,0.000){0}{\usebox{\plotpoint}}
\put(286.04,139.00){\usebox{\plotpoint}}
\put(306.80,139.00){\usebox{\plotpoint}}
\multiput(309,139)(20.756,0.000){0}{\usebox{\plotpoint}}
\put(327.55,139.00){\usebox{\plotpoint}}
\put(348.31,139.00){\usebox{\plotpoint}}
\multiput(349,139)(20.756,0.000){0}{\usebox{\plotpoint}}
\put(369.07,139.00){\usebox{\plotpoint}}
\put(389.82,139.00){\usebox{\plotpoint}}
\multiput(390,139)(20.756,0.000){0}{\usebox{\plotpoint}}
\put(410.58,139.00){\usebox{\plotpoint}}
\multiput(417,139)(20.756,0.000){0}{\usebox{\plotpoint}}
\put(431.33,139.00){\usebox{\plotpoint}}
\put(452.09,139.00){\usebox{\plotpoint}}
\multiput(457,139)(20.756,0.000){0}{\usebox{\plotpoint}}
\put(472.84,139.00){\usebox{\plotpoint}}
\put(493.60,139.00){\usebox{\plotpoint}}
\multiput(498,139)(20.756,0.000){0}{\usebox{\plotpoint}}
\put(514.35,139.00){\usebox{\plotpoint}}
\put(535.11,139.00){\usebox{\plotpoint}}
\multiput(538,139)(20.756,0.000){0}{\usebox{\plotpoint}}
\put(555.87,139.00){\usebox{\plotpoint}}
\put(576.62,139.00){\usebox{\plotpoint}}
\multiput(579,139)(20.756,0.000){0}{\usebox{\plotpoint}}
\put(597.38,139.00){\usebox{\plotpoint}}
\put(618.13,139.00){\usebox{\plotpoint}}
\multiput(619,139)(20.756,0.000){0}{\usebox{\plotpoint}}
\put(638.89,139.00){\usebox{\plotpoint}}
\put(659.64,139.00){\usebox{\plotpoint}}
\multiput(660,139)(20.756,0.000){0}{\usebox{\plotpoint}}
\put(680.40,139.00){\usebox{\plotpoint}}
\multiput(687,139)(20.756,0.000){0}{\usebox{\plotpoint}}
\put(701.15,139.00){\usebox{\plotpoint}}
\put(721.91,139.00){\usebox{\plotpoint}}
\multiput(727,139)(20.756,0.000){0}{\usebox{\plotpoint}}
\put(742.66,139.00){\usebox{\plotpoint}}
\put(763.42,139.00){\usebox{\plotpoint}}
\multiput(768,139)(20.756,0.000){0}{\usebox{\plotpoint}}
\put(784.18,139.00){\usebox{\plotpoint}}
\put(804.93,139.00){\usebox{\plotpoint}}
\multiput(808,139)(20.756,0.000){0}{\usebox{\plotpoint}}
\put(825.69,139.00){\usebox{\plotpoint}}
\put(846.44,139.00){\usebox{\plotpoint}}
\multiput(849,139)(20.756,0.000){0}{\usebox{\plotpoint}}
\put(867.20,139.00){\usebox{\plotpoint}}
\put(887.95,139.00){\usebox{\plotpoint}}
\multiput(889,139)(20.756,0.000){0}{\usebox{\plotpoint}}
\put(908.71,139.00){\usebox{\plotpoint}}
\put(929.46,139.00){\usebox{\plotpoint}}
\multiput(930,139)(20.756,0.000){0}{\usebox{\plotpoint}}
\put(950.22,139.00){\usebox{\plotpoint}}
\multiput(957,139)(20.756,0.000){0}{\usebox{\plotpoint}}
\put(970.98,139.00){\usebox{\plotpoint}}
\put(991.73,139.00){\usebox{\plotpoint}}
\multiput(997,139)(20.756,0.000){0}{\usebox{\plotpoint}}
\put(1012.49,139.00){\usebox{\plotpoint}}
\put(1033.24,139.00){\usebox{\plotpoint}}
\multiput(1038,139)(20.756,0.000){0}{\usebox{\plotpoint}}
\put(1054.00,139.00){\usebox{\plotpoint}}
\put(1074.75,139.00){\usebox{\plotpoint}}
\multiput(1078,139)(20.756,0.000){0}{\usebox{\plotpoint}}
\put(1095.51,139.00){\usebox{\plotpoint}}
\put(1116.26,139.00){\usebox{\plotpoint}}
\multiput(1119,139)(20.756,0.000){0}{\usebox{\plotpoint}}
\put(1137.02,139.00){\usebox{\plotpoint}}
\put(1157.77,139.00){\usebox{\plotpoint}}
\multiput(1159,139)(20.756,0.000){0}{\usebox{\plotpoint}}
\put(1178.53,139.00){\usebox{\plotpoint}}
\put(1199.29,139.00){\usebox{\plotpoint}}
\multiput(1200,139)(20.756,0.000){0}{\usebox{\plotpoint}}
\put(1220.04,139.00){\usebox{\plotpoint}}
\multiput(1227,139)(20.756,0.000){0}{\usebox{\plotpoint}}
\put(1240.80,139.00){\usebox{\plotpoint}}
\put(1261.55,139.00){\usebox{\plotpoint}}
\multiput(1267,139)(20.756,0.000){0}{\usebox{\plotpoint}}
\put(1282.31,139.00){\usebox{\plotpoint}}
\put(1303.06,139.00){\usebox{\plotpoint}}
\multiput(1308,139)(20.756,0.000){0}{\usebox{\plotpoint}}
\put(1323.82,139.00){\usebox{\plotpoint}}
\put(1344.57,139.00){\usebox{\plotpoint}}
\multiput(1348,139)(20.756,0.000){0}{\usebox{\plotpoint}}
\put(1365.33,139.00){\usebox{\plotpoint}}
\put(1386.09,139.00){\usebox{\plotpoint}}
\multiput(1389,139)(20.756,0.000){0}{\usebox{\plotpoint}}
\put(1406.84,139.00){\usebox{\plotpoint}}
\put(1427.60,139.00){\usebox{\plotpoint}}
\multiput(1429,139)(20.756,0.000){0}{\usebox{\plotpoint}}
\put(1448.35,139.00){\usebox{\plotpoint}}
\put(1456,139){\usebox{\plotpoint}}
\end{picture}

%% file: b23_du1.tex
\setlength{\unitlength}{0.240900pt}
\ifx\plotpoint\undefined\newsavebox{\plotpoint}\fi
\begin{picture}(1500,900)(0,0)
\font\gnuplot=cmr10 at 12pt
\gnuplot
\sbox{\plotpoint}{\rule[-0.200pt]{0.400pt}{0.400pt}}%
\put(120.0,103.0){\rule[-0.200pt]{321.842pt}{0.400pt}}
\put(120.0,103.0){\rule[-0.200pt]{4.818pt}{0.400pt}}
\put(108,103){\makebox(0,0)[r]{$0$}}
\put(1436.0,103.0){\rule[-0.200pt]{4.818pt}{0.400pt}}
\put(120.0,282.0){\rule[-0.200pt]{4.818pt}{0.400pt}}
\put(108,282){\makebox(0,0)[r]{$0.05$}}
\put(1436.0,282.0){\rule[-0.200pt]{4.818pt}{0.400pt}}
\put(120.0,462.0){\rule[-0.200pt]{4.818pt}{0.400pt}}
\put(108,462){\makebox(0,0)[r]{$0.1$}}
\put(1436.0,462.0){\rule[-0.200pt]{4.818pt}{0.400pt}}
\put(120.0,642.0){\rule[-0.200pt]{4.818pt}{0.400pt}}
\put(108,642){\makebox(0,0)[r]{$0.15$}}
\put(1436.0,642.0){\rule[-0.200pt]{4.818pt}{0.400pt}}
\put(120.0,821.0){\rule[-0.200pt]{4.818pt}{0.400pt}}
\put(108,821){\makebox(0,0)[r]{$0.2$}}
\put(1436.0,821.0){\rule[-0.200pt]{4.818pt}{0.400pt}}
\put(236.0,31.0){\rule[-0.200pt]{0.400pt}{4.818pt}}
\put(236,19){\makebox(0,0){\shortstack{\\ \\ \\ \\ $5$}}}
\put(236.0,873.0){\rule[-0.200pt]{0.400pt}{4.818pt}}
\put(527.0,31.0){\rule[-0.200pt]{0.400pt}{4.818pt}}
\put(527,19){\makebox(0,0){\shortstack{\\ \\ \\ \\ $10$}}}
\put(527.0,873.0){\rule[-0.200pt]{0.400pt}{4.818pt}}
\put(817.0,31.0){\rule[-0.200pt]{0.400pt}{4.818pt}}
\put(817,19){\makebox(0,0){\shortstack{\\ \\ \\ \\ $15$}}}
\put(817.0,873.0){\rule[-0.200pt]{0.400pt}{4.818pt}}
\put(1107.0,31.0){\rule[-0.200pt]{0.400pt}{4.818pt}}
\put(1107,19){\makebox(0,0){\shortstack{\\ \\ \\ \\ $20$}}}
\put(1107.0,873.0){\rule[-0.200pt]{0.400pt}{4.818pt}}
\put(1398.0,31.0){\rule[-0.200pt]{0.400pt}{4.818pt}}
\put(1398,19){\makebox(0,0){\shortstack{\\ \\ \\ \\ $25$}}}
\put(1398.0,873.0){\rule[-0.200pt]{0.400pt}{4.818pt}}
\put(120.0,31.0){\rule[-0.200pt]{321.842pt}{0.400pt}}
\put(1456.0,31.0){\rule[-0.200pt]{0.400pt}{207.656pt}}
\put(120.0,893.0){\rule[-0.200pt]{321.842pt}{0.400pt}}
\put(12,534){\makebox(0,0){$\sigma_{\hbox{\small eff}} (r,t)$}}
\put(788,-53){\makebox(0,0){$\hbox{Area} = r \times t$}}
\put(120.0,31.0){\rule[-0.200pt]{0.400pt}{207.656pt}}
\put(178,519){\makebox(0,0){$\times$}}
\put(288,351){\makebox(0,0){$\times$}}
\put(300,369){\makebox(0,0){$\times$}}
\put(405,326){\makebox(0,0){$\times$}}
\put(416,322){\makebox(0,0){$\times$}}
\put(469,121){\makebox(0,0){$\times$}}
\put(521,221){\makebox(0,0){$\times$}}
\put(532,282){\makebox(0,0){$\times$}}
\put(637,282){\makebox(0,0){$\times$}}
\put(649,214){\makebox(0,0){$\times$}}
\put(811,114){\makebox(0,0){$\times$}}
\put(823,229){\makebox(0,0){$\times$}}
\put(875,167){\makebox(0,0){$\times$}}
\put(985,390){\makebox(0,0){$\times$}}
\put(997,81){\makebox(0,0){$\times$}}
\put(1102,171){\makebox(0,0){$\times$}}
\put(1113,189){\makebox(0,0){$\times$}}
\put(1398,300){\makebox(0,0){$\times$}}
\put(178.0,505.0){\rule[-0.200pt]{0.400pt}{6.986pt}}
\put(168.0,505.0){\rule[-0.200pt]{4.818pt}{0.400pt}}
\put(168.0,534.0){\rule[-0.200pt]{4.818pt}{0.400pt}}
\put(288.0,326.0){\rule[-0.200pt]{0.400pt}{12.045pt}}
\put(278.0,326.0){\rule[-0.200pt]{4.818pt}{0.400pt}}
\put(278.0,376.0){\rule[-0.200pt]{4.818pt}{0.400pt}}
\put(300.0,343.0){\rule[-0.200pt]{0.400pt}{12.286pt}}
\put(290.0,343.0){\rule[-0.200pt]{4.818pt}{0.400pt}}
\put(290.0,394.0){\rule[-0.200pt]{4.818pt}{0.400pt}}
\put(405.0,300.0){\rule[-0.200pt]{0.400pt}{12.286pt}}
\put(395.0,300.0){\rule[-0.200pt]{4.818pt}{0.400pt}}
\put(395.0,351.0){\rule[-0.200pt]{4.818pt}{0.400pt}}
\put(416.0,286.0){\rule[-0.200pt]{0.400pt}{17.345pt}}
\put(406.0,286.0){\rule[-0.200pt]{4.818pt}{0.400pt}}
\put(406.0,358.0){\rule[-0.200pt]{4.818pt}{0.400pt}}
\put(469.0,85.0){\rule[-0.200pt]{0.400pt}{17.345pt}}
\put(459.0,85.0){\rule[-0.200pt]{4.818pt}{0.400pt}}
\put(459.0,157.0){\rule[-0.200pt]{4.818pt}{0.400pt}}
\put(521.0,167.0){\rule[-0.200pt]{0.400pt}{26.017pt}}
\put(511.0,167.0){\rule[-0.200pt]{4.818pt}{0.400pt}}
\put(511.0,275.0){\rule[-0.200pt]{4.818pt}{0.400pt}}
\put(532.0,236.0){\rule[-0.200pt]{0.400pt}{22.404pt}}
\put(522.0,236.0){\rule[-0.200pt]{4.818pt}{0.400pt}}
\put(522.0,329.0){\rule[-0.200pt]{4.818pt}{0.400pt}}
\put(637.0,221.0){\rule[-0.200pt]{0.400pt}{29.390pt}}
\put(627.0,221.0){\rule[-0.200pt]{4.818pt}{0.400pt}}
\put(627.0,343.0){\rule[-0.200pt]{4.818pt}{0.400pt}}
\put(649.0,146.0){\rule[-0.200pt]{0.400pt}{32.762pt}}
\put(639.0,146.0){\rule[-0.200pt]{4.818pt}{0.400pt}}
\put(639.0,282.0){\rule[-0.200pt]{4.818pt}{0.400pt}}
\put(811.0,35.0){\rule[-0.200pt]{0.400pt}{38.062pt}}
\put(801.0,35.0){\rule[-0.200pt]{4.818pt}{0.400pt}}
\put(801.0,193.0){\rule[-0.200pt]{4.818pt}{0.400pt}}
\put(823.0,132.0){\rule[-0.200pt]{0.400pt}{46.735pt}}
\put(813.0,132.0){\rule[-0.200pt]{4.818pt}{0.400pt}}
\put(813.0,326.0){\rule[-0.200pt]{4.818pt}{0.400pt}}
\put(875.0,60.0){\rule[-0.200pt]{0.400pt}{51.793pt}}
\put(865.0,60.0){\rule[-0.200pt]{4.818pt}{0.400pt}}
\put(865.0,275.0){\rule[-0.200pt]{4.818pt}{0.400pt}}
\put(985.0,218.0){\rule[-0.200pt]{0.400pt}{83.110pt}}
\put(975.0,218.0){\rule[-0.200pt]{4.818pt}{0.400pt}}
\put(975.0,563.0){\rule[-0.200pt]{4.818pt}{0.400pt}}
\put(997.0,31.0){\rule[-0.200pt]{0.400pt}{44.085pt}}
\put(987.0,31.0){\rule[-0.200pt]{4.818pt}{0.400pt}}
\put(987.0,214.0){\rule[-0.200pt]{4.818pt}{0.400pt}}
\put(1102.0,31.0){\rule[-0.200pt]{0.400pt}{69.138pt}}
\put(1092.0,31.0){\rule[-0.200pt]{4.818pt}{0.400pt}}
\put(1092.0,318.0){\rule[-0.200pt]{4.818pt}{0.400pt}}
\put(1113.0,31.0){\rule[-0.200pt]{0.400pt}{79.497pt}}
\put(1103.0,31.0){\rule[-0.200pt]{4.818pt}{0.400pt}}
\put(1103.0,361.0){\rule[-0.200pt]{4.818pt}{0.400pt}}
\put(1398.0,42.0){\rule[-0.200pt]{0.400pt}{124.545pt}}
\put(1388.0,42.0){\rule[-0.200pt]{4.818pt}{0.400pt}}
\put(1388.0,559.0){\rule[-0.200pt]{4.818pt}{0.400pt}}
\end{picture}

%% file: b24_du1.tex
\setlength{\unitlength}{0.240900pt}
\ifx\plotpoint\undefined\newsavebox{\plotpoint}\fi
\begin{picture}(1500,900)(0,0)
\font\gnuplot=cmr10 at 12pt
\gnuplot
\sbox{\plotpoint}{\rule[-0.200pt]{0.400pt}{0.400pt}}%
\put(120.0,146.0){\rule[-0.200pt]{321.842pt}{0.400pt}}
\put(120.0,31.0){\rule[-0.200pt]{4.818pt}{0.400pt}}
\put(108,31){\makebox(0,0)[r]{$-0.02$}}
\put(1436.0,31.0){\rule[-0.200pt]{4.818pt}{0.400pt}}
\put(120.0,146.0){\rule[-0.200pt]{4.818pt}{0.400pt}}
\put(108,146){\makebox(0,0)[r]{$0$}}
\put(1436.0,146.0){\rule[-0.200pt]{4.818pt}{0.400pt}}
\put(120.0,261.0){\rule[-0.200pt]{4.818pt}{0.400pt}}
\put(108,261){\makebox(0,0)[r]{$0.02$}}
\put(1436.0,261.0){\rule[-0.200pt]{4.818pt}{0.400pt}}
\put(120.0,376.0){\rule[-0.200pt]{4.818pt}{0.400pt}}
\put(108,376){\makebox(0,0)[r]{$0.04$}}
\put(1436.0,376.0){\rule[-0.200pt]{4.818pt}{0.400pt}}
\put(120.0,491.0){\rule[-0.200pt]{4.818pt}{0.400pt}}
\put(108,491){\makebox(0,0)[r]{$0.06$}}
\put(1436.0,491.0){\rule[-0.200pt]{4.818pt}{0.400pt}}
\put(120.0,606.0){\rule[-0.200pt]{4.818pt}{0.400pt}}
\put(108,606){\makebox(0,0)[r]{$0.08$}}
\put(1436.0,606.0){\rule[-0.200pt]{4.818pt}{0.400pt}}
\put(120.0,721.0){\rule[-0.200pt]{4.818pt}{0.400pt}}
\put(108,721){\makebox(0,0)[r]{$0.1$}}
\put(1436.0,721.0){\rule[-0.200pt]{4.818pt}{0.400pt}}
\put(120.0,836.0){\rule[-0.200pt]{4.818pt}{0.400pt}}
\put(108,836){\makebox(0,0)[r]{$0.12$}}
\put(1436.0,836.0){\rule[-0.200pt]{4.818pt}{0.400pt}}
\put(236.0,31.0){\rule[-0.200pt]{0.400pt}{4.818pt}}
\put(236,19){\makebox(0,0){\shortstack{\\ \\ \\ \\ $5$}}}
\put(236.0,873.0){\rule[-0.200pt]{0.400pt}{4.818pt}}
\put(527.0,31.0){\rule[-0.200pt]{0.400pt}{4.818pt}}
\put(527,19){\makebox(0,0){\shortstack{\\ \\ \\ \\ $10$}}}
\put(527.0,873.0){\rule[-0.200pt]{0.400pt}{4.818pt}}
\put(817.0,31.0){\rule[-0.200pt]{0.400pt}{4.818pt}}
\put(817,19){\makebox(0,0){\shortstack{\\ \\ \\ \\ $15$}}}
\put(817.0,873.0){\rule[-0.200pt]{0.400pt}{4.818pt}}
\put(1107.0,31.0){\rule[-0.200pt]{0.400pt}{4.818pt}}
\put(1107,19){\makebox(0,0){\shortstack{\\ \\ \\ \\ $20$}}}
\put(1107.0,873.0){\rule[-0.200pt]{0.400pt}{4.818pt}}
\put(1398.0,31.0){\rule[-0.200pt]{0.400pt}{4.818pt}}
\put(1398,19){\makebox(0,0){\shortstack{\\ \\ \\ \\ $25$}}}
\put(1398.0,873.0){\rule[-0.200pt]{0.400pt}{4.818pt}}
\put(120.0,31.0){\rule[-0.200pt]{321.842pt}{0.400pt}}
\put(1456.0,31.0){\rule[-0.200pt]{0.400pt}{207.656pt}}
\put(120.0,893.0){\rule[-0.200pt]{321.842pt}{0.400pt}}
\put(12,438){\makebox(0,0){$\sigma_{eff} (r,t)$}}
\put(788,-53){\makebox(0,0){$\hbox{Area} = r \times t$}}
\put(120.0,31.0){\rule[-0.200pt]{0.400pt}{207.656pt}}
\put(178,778){\makebox(0,0){$\times$}}
\put(288,548){\makebox(0,0){$\times$}}
\put(300,473){\makebox(0,0){$\times$}}
\put(405,393){\makebox(0,0){$\times$}}
\put(416,496){\makebox(0,0){$\times$}}
\put(469,370){\makebox(0,0){$\times$}}
\put(521,387){\makebox(0,0){$\times$}}
\put(532,370){\makebox(0,0){$\times$}}
\put(637,267){\makebox(0,0){$\times$}}
\put(649,129){\makebox(0,0){$\times$}}
\put(811,284){\makebox(0,0){$\times$}}
\put(823,376){\makebox(0,0){$\times$}}
\put(875,514){\makebox(0,0){$\times$}}
\put(985,226){\makebox(0,0){$\times$}}
\put(997,106){\makebox(0,0){$\times$}}
\put(1398,502){\makebox(0,0){$\times$}}
\put(178.0,744.0){\rule[-0.200pt]{0.400pt}{16.622pt}}
\put(168.0,744.0){\rule[-0.200pt]{4.818pt}{0.400pt}}
\put(168.0,813.0){\rule[-0.200pt]{4.818pt}{0.400pt}}
\put(288.0,508.0){\rule[-0.200pt]{0.400pt}{19.272pt}}
\put(278.0,508.0){\rule[-0.200pt]{4.818pt}{0.400pt}}
\put(278.0,588.0){\rule[-0.200pt]{4.818pt}{0.400pt}}
\put(300.0,445.0){\rule[-0.200pt]{0.400pt}{13.731pt}}
\put(290.0,445.0){\rule[-0.200pt]{4.818pt}{0.400pt}}
\put(290.0,502.0){\rule[-0.200pt]{4.818pt}{0.400pt}}
\put(405.0,347.0){\rule[-0.200pt]{0.400pt}{22.163pt}}
\put(395.0,347.0){\rule[-0.200pt]{4.818pt}{0.400pt}}
\put(395.0,439.0){\rule[-0.200pt]{4.818pt}{0.400pt}}
\put(416.0,451.0){\rule[-0.200pt]{0.400pt}{21.922pt}}
\put(406.0,451.0){\rule[-0.200pt]{4.818pt}{0.400pt}}
\put(406.0,542.0){\rule[-0.200pt]{4.818pt}{0.400pt}}
\put(469.0,290.0){\rule[-0.200pt]{0.400pt}{38.785pt}}
\put(459.0,290.0){\rule[-0.200pt]{4.818pt}{0.400pt}}
\put(459.0,451.0){\rule[-0.200pt]{4.818pt}{0.400pt}}
\put(521.0,307.0){\rule[-0.200pt]{0.400pt}{38.785pt}}
\put(511.0,307.0){\rule[-0.200pt]{4.818pt}{0.400pt}}
\put(511.0,468.0){\rule[-0.200pt]{4.818pt}{0.400pt}}
\put(532.0,301.0){\rule[-0.200pt]{0.400pt}{33.244pt}}
\put(522.0,301.0){\rule[-0.200pt]{4.818pt}{0.400pt}}
\put(522.0,439.0){\rule[-0.200pt]{4.818pt}{0.400pt}}
\put(637.0,180.0){\rule[-0.200pt]{0.400pt}{41.676pt}}
\put(627.0,180.0){\rule[-0.200pt]{4.818pt}{0.400pt}}
\put(627.0,353.0){\rule[-0.200pt]{4.818pt}{0.400pt}}
\put(649.0,48.0){\rule[-0.200pt]{0.400pt}{38.785pt}}
\put(639.0,48.0){\rule[-0.200pt]{4.818pt}{0.400pt}}
\put(639.0,209.0){\rule[-0.200pt]{4.818pt}{0.400pt}}
\put(811.0,157.0){\rule[-0.200pt]{0.400pt}{60.948pt}}
\put(801.0,157.0){\rule[-0.200pt]{4.818pt}{0.400pt}}
\put(801.0,410.0){\rule[-0.200pt]{4.818pt}{0.400pt}}
\put(823.0,284.0){\rule[-0.200pt]{0.400pt}{44.326pt}}
\put(813.0,284.0){\rule[-0.200pt]{4.818pt}{0.400pt}}
\put(813.0,468.0){\rule[-0.200pt]{4.818pt}{0.400pt}}
\put(875.0,376.0){\rule[-0.200pt]{0.400pt}{66.488pt}}
\put(865.0,376.0){\rule[-0.200pt]{4.818pt}{0.400pt}}
\put(865.0,652.0){\rule[-0.200pt]{4.818pt}{0.400pt}}
\put(985.0,83.0){\rule[-0.200pt]{0.400pt}{69.138pt}}
\put(975.0,83.0){\rule[-0.200pt]{4.818pt}{0.400pt}}
\put(975.0,370.0){\rule[-0.200pt]{4.818pt}{0.400pt}}
\put(997.0,31.0){\rule[-0.200pt]{0.400pt}{49.866pt}}
\put(987.0,31.0){\rule[-0.200pt]{4.818pt}{0.400pt}}
\put(987.0,238.0){\rule[-0.200pt]{4.818pt}{0.400pt}}
\put(1102.0,31.0){\rule[-0.200pt]{0.400pt}{33.244pt}}
\put(1092.0,31.0){\rule[-0.200pt]{4.818pt}{0.400pt}}
\put(1092.0,169.0){\rule[-0.200pt]{4.818pt}{0.400pt}}
\put(1398.0,232.0){\rule[-0.200pt]{0.400pt}{130.086pt}}
\put(1388.0,232.0){\rule[-0.200pt]{4.818pt}{0.400pt}}
\put(1388.0,772.0){\rule[-0.200pt]{4.818pt}{0.400pt}}
\end{picture}

%% file: b23_dgtd.tex
\setlength{\unitlength}{0.240900pt}
\ifx\plotpoint\undefined\newsavebox{\plotpoint}\fi
\begin{picture}(1500,900)(0,0)
\font\gnuplot=cmr10 at 12pt
\gnuplot
\sbox{\plotpoint}{\rule[-0.200pt]{0.400pt}{0.400pt}}%
\put(120.0,103.0){\rule[-0.200pt]{321.842pt}{0.400pt}}
\put(120.0,103.0){\rule[-0.200pt]{4.818pt}{0.400pt}}
\put(108,103){\makebox(0,0)[r]{$0$}}
\put(1436.0,103.0){\rule[-0.200pt]{4.818pt}{0.400pt}}
\put(120.0,282.0){\rule[-0.200pt]{4.818pt}{0.400pt}}
\put(108,282){\makebox(0,0)[r]{$0.05$}}
\put(1436.0,282.0){\rule[-0.200pt]{4.818pt}{0.400pt}}
\put(120.0,462.0){\rule[-0.200pt]{4.818pt}{0.400pt}}
\put(108,462){\makebox(0,0)[r]{$0.1$}}
\put(1436.0,462.0){\rule[-0.200pt]{4.818pt}{0.400pt}}
\put(120.0,642.0){\rule[-0.200pt]{4.818pt}{0.400pt}}
\put(108,642){\makebox(0,0)[r]{$0.15$}}
\put(1436.0,642.0){\rule[-0.200pt]{4.818pt}{0.400pt}}
\put(120.0,821.0){\rule[-0.200pt]{4.818pt}{0.400pt}}
\put(108,821){\makebox(0,0)[r]{$0.2$}}
\put(1436.0,821.0){\rule[-0.200pt]{4.818pt}{0.400pt}}
\put(236.0,31.0){\rule[-0.200pt]{0.400pt}{4.818pt}}
\put(236,19){\makebox(0,0){\shortstack{\\ \\ \\ \\ $5$}}}
\put(236.0,873.0){\rule[-0.200pt]{0.400pt}{4.818pt}}
\put(527.0,31.0){\rule[-0.200pt]{0.400pt}{4.818pt}}
\put(527,19){\makebox(0,0){\shortstack{\\ \\ \\ \\ $10$}}}
\put(527.0,873.0){\rule[-0.200pt]{0.400pt}{4.818pt}}
\put(817.0,31.0){\rule[-0.200pt]{0.400pt}{4.818pt}}
\put(817,19){\makebox(0,0){\shortstack{\\ \\ \\ \\ $15$}}}
\put(817.0,873.0){\rule[-0.200pt]{0.400pt}{4.818pt}}
\put(1107.0,31.0){\rule[-0.200pt]{0.400pt}{4.818pt}}
\put(1107,19){\makebox(0,0){\shortstack{\\ \\ \\ \\ $20$}}}
\put(1107.0,873.0){\rule[-0.200pt]{0.400pt}{4.818pt}}
\put(1398.0,31.0){\rule[-0.200pt]{0.400pt}{4.818pt}}
\put(1398,19){\makebox(0,0){\shortstack{\\ \\ \\ \\ $25$}}}
\put(1398.0,873.0){\rule[-0.200pt]{0.400pt}{4.818pt}}
\put(120.0,31.0){\rule[-0.200pt]{321.842pt}{0.400pt}}
\put(1456.0,31.0){\rule[-0.200pt]{0.400pt}{207.656pt}}
\put(120.0,893.0){\rule[-0.200pt]{321.842pt}{0.400pt}}
\put(12,534){\makebox(0,0){$\sigma_{\hbox{\small eff}} (r,t)$}}
\put(788,-53){\makebox(0,0){$\hbox{Area} = r \times t$}}
\put(120.0,31.0){\rule[-0.200pt]{0.400pt}{207.656pt}}
\put(178,573){\makebox(0,0){$+$}}
\put(288,584){\makebox(0,0){$+$}}
\put(300,581){\makebox(0,0){$+$}}
\put(405,588){\makebox(0,0){$+$}}
\put(416,584){\makebox(0,0){$+$}}
\put(469,584){\makebox(0,0){$+$}}
\put(521,581){\makebox(0,0){$+$}}
\put(532,588){\makebox(0,0){$+$}}
\put(637,570){\makebox(0,0){$+$}}
\put(649,588){\makebox(0,0){$+$}}
\put(811,559){\makebox(0,0){$+$}}
\put(823,591){\makebox(0,0){$+$}}
\put(875,570){\makebox(0,0){$+$}}
\put(985,573){\makebox(0,0){$+$}}
\put(997,584){\makebox(0,0){$+$}}
\put(1102,552){\makebox(0,0){$+$}}
\put(1113,537){\makebox(0,0){$+$}}
\put(1398,527){\makebox(0,0){$+$}}
\put(178.0,570.0){\rule[-0.200pt]{0.400pt}{1.686pt}}
\put(168.0,570.0){\rule[-0.200pt]{4.818pt}{0.400pt}}
\put(168.0,577.0){\rule[-0.200pt]{4.818pt}{0.400pt}}
\put(288.0,577.0){\rule[-0.200pt]{0.400pt}{3.373pt}}
\put(278.0,577.0){\rule[-0.200pt]{4.818pt}{0.400pt}}
\put(278.0,591.0){\rule[-0.200pt]{4.818pt}{0.400pt}}
\put(300.0,577.0){\rule[-0.200pt]{0.400pt}{1.686pt}}
\put(290.0,577.0){\rule[-0.200pt]{4.818pt}{0.400pt}}
\put(290.0,584.0){\rule[-0.200pt]{4.818pt}{0.400pt}}
\put(405.0,584.0){\rule[-0.200pt]{0.400pt}{1.686pt}}
\put(395.0,584.0){\rule[-0.200pt]{4.818pt}{0.400pt}}
\put(395.0,591.0){\rule[-0.200pt]{4.818pt}{0.400pt}}
\put(416.0,577.0){\rule[-0.200pt]{0.400pt}{3.373pt}}
\put(406.0,577.0){\rule[-0.200pt]{4.818pt}{0.400pt}}
\put(406.0,591.0){\rule[-0.200pt]{4.818pt}{0.400pt}}
\put(469.0,577.0){\rule[-0.200pt]{0.400pt}{3.373pt}}
\put(459.0,577.0){\rule[-0.200pt]{4.818pt}{0.400pt}}
\put(459.0,591.0){\rule[-0.200pt]{4.818pt}{0.400pt}}
\put(521.0,573.0){\rule[-0.200pt]{0.400pt}{3.613pt}}
\put(511.0,573.0){\rule[-0.200pt]{4.818pt}{0.400pt}}
\put(511.0,588.0){\rule[-0.200pt]{4.818pt}{0.400pt}}
\put(532.0,581.0){\rule[-0.200pt]{0.400pt}{3.373pt}}
\put(522.0,581.0){\rule[-0.200pt]{4.818pt}{0.400pt}}
\put(522.0,595.0){\rule[-0.200pt]{4.818pt}{0.400pt}}
\put(637.0,563.0){\rule[-0.200pt]{0.400pt}{3.373pt}}
\put(627.0,563.0){\rule[-0.200pt]{4.818pt}{0.400pt}}
\put(627.0,577.0){\rule[-0.200pt]{4.818pt}{0.400pt}}
\put(649.0,581.0){\rule[-0.200pt]{0.400pt}{3.373pt}}
\put(639.0,581.0){\rule[-0.200pt]{4.818pt}{0.400pt}}
\put(639.0,595.0){\rule[-0.200pt]{4.818pt}{0.400pt}}
\put(811.0,545.0){\rule[-0.200pt]{0.400pt}{6.745pt}}
\put(801.0,545.0){\rule[-0.200pt]{4.818pt}{0.400pt}}
\put(801.0,573.0){\rule[-0.200pt]{4.818pt}{0.400pt}}
\put(823.0,577.0){\rule[-0.200pt]{0.400pt}{6.986pt}}
\put(813.0,577.0){\rule[-0.200pt]{4.818pt}{0.400pt}}
\put(813.0,606.0){\rule[-0.200pt]{4.818pt}{0.400pt}}
\put(875.0,552.0){\rule[-0.200pt]{0.400pt}{8.672pt}}
\put(865.0,552.0){\rule[-0.200pt]{4.818pt}{0.400pt}}
\put(865.0,588.0){\rule[-0.200pt]{4.818pt}{0.400pt}}
\put(985.0,552.0){\rule[-0.200pt]{0.400pt}{10.359pt}}
\put(975.0,552.0){\rule[-0.200pt]{4.818pt}{0.400pt}}
\put(975.0,595.0){\rule[-0.200pt]{4.818pt}{0.400pt}}
\put(997.0,566.0){\rule[-0.200pt]{0.400pt}{8.672pt}}
\put(987.0,566.0){\rule[-0.200pt]{4.818pt}{0.400pt}}
\put(987.0,602.0){\rule[-0.200pt]{4.818pt}{0.400pt}}
\put(1102.0,523.0){\rule[-0.200pt]{0.400pt}{13.972pt}}
\put(1092.0,523.0){\rule[-0.200pt]{4.818pt}{0.400pt}}
\put(1092.0,581.0){\rule[-0.200pt]{4.818pt}{0.400pt}}
\put(1113.0,512.0){\rule[-0.200pt]{0.400pt}{12.286pt}}
\put(1103.0,512.0){\rule[-0.200pt]{4.818pt}{0.400pt}}
\put(1103.0,563.0){\rule[-0.200pt]{4.818pt}{0.400pt}}
\put(1398.0,466.0){\rule[-0.200pt]{0.400pt}{29.390pt}}
\put(1388.0,466.0){\rule[-0.200pt]{4.818pt}{0.400pt}}
\put(1388.0,588.0){\rule[-0.200pt]{4.818pt}{0.400pt}}
\put(178,810){\makebox(0,0){$\times$}}
\put(288,663){\makebox(0,0){$\times$}}
\put(300,670){\makebox(0,0){$\times$}}
\put(405,581){\makebox(0,0){$\times$}}
\put(416,602){\makebox(0,0){$\times$}}
\put(469,455){\makebox(0,0){$\times$}}
\put(521,545){\makebox(0,0){$\times$}}
\put(532,534){\makebox(0,0){$\times$}}
\put(637,376){\makebox(0,0){$\times$}}
\put(649,394){\makebox(0,0){$\times$}}
\put(811,315){\makebox(0,0){$\times$}}
\put(823,322){\makebox(0,0){$\times$}}
\put(875,272){\makebox(0,0){$\times$}}
\put(985,293){\makebox(0,0){$\times$}}
\put(997,340){\makebox(0,0){$\times$}}
\put(1102,189){\makebox(0,0){$\times$}}
\put(1113,361){\makebox(0,0){$\times$}}
\put(1398,268){\makebox(0,0){$\times$}}
\put(178.0,803.0){\rule[-0.200pt]{0.400pt}{3.613pt}}
\put(168.0,803.0){\rule[-0.200pt]{4.818pt}{0.400pt}}
\put(168.0,818.0){\rule[-0.200pt]{4.818pt}{0.400pt}}
\put(288.0,652.0){\rule[-0.200pt]{0.400pt}{5.300pt}}
\put(278.0,652.0){\rule[-0.200pt]{4.818pt}{0.400pt}}
\put(278.0,674.0){\rule[-0.200pt]{4.818pt}{0.400pt}}
\put(300.0,660.0){\rule[-0.200pt]{0.400pt}{5.059pt}}
\put(290.0,660.0){\rule[-0.200pt]{4.818pt}{0.400pt}}
\put(290.0,681.0){\rule[-0.200pt]{4.818pt}{0.400pt}}
\put(405.0,566.0){\rule[-0.200pt]{0.400pt}{6.986pt}}
\put(395.0,566.0){\rule[-0.200pt]{4.818pt}{0.400pt}}
\put(395.0,595.0){\rule[-0.200pt]{4.818pt}{0.400pt}}
\put(416.0,591.0){\rule[-0.200pt]{0.400pt}{5.300pt}}
\put(406.0,591.0){\rule[-0.200pt]{4.818pt}{0.400pt}}
\put(406.0,613.0){\rule[-0.200pt]{4.818pt}{0.400pt}}
\put(469.0,437.0){\rule[-0.200pt]{0.400pt}{8.672pt}}
\put(459.0,437.0){\rule[-0.200pt]{4.818pt}{0.400pt}}
\put(459.0,473.0){\rule[-0.200pt]{4.818pt}{0.400pt}}
\put(521.0,527.0){\rule[-0.200pt]{0.400pt}{8.672pt}}
\put(511.0,527.0){\rule[-0.200pt]{4.818pt}{0.400pt}}
\put(511.0,563.0){\rule[-0.200pt]{4.818pt}{0.400pt}}
\put(532.0,516.0){\rule[-0.200pt]{0.400pt}{8.672pt}}
\put(522.0,516.0){\rule[-0.200pt]{4.818pt}{0.400pt}}
\put(522.0,552.0){\rule[-0.200pt]{4.818pt}{0.400pt}}
\put(637.0,354.0){\rule[-0.200pt]{0.400pt}{10.359pt}}
\put(627.0,354.0){\rule[-0.200pt]{4.818pt}{0.400pt}}
\put(627.0,397.0){\rule[-0.200pt]{4.818pt}{0.400pt}}
\put(649.0,369.0){\rule[-0.200pt]{0.400pt}{12.045pt}}
\put(639.0,369.0){\rule[-0.200pt]{4.818pt}{0.400pt}}
\put(639.0,419.0){\rule[-0.200pt]{4.818pt}{0.400pt}}
\put(811.0,279.0){\rule[-0.200pt]{0.400pt}{17.345pt}}
\put(801.0,279.0){\rule[-0.200pt]{4.818pt}{0.400pt}}
\put(801.0,351.0){\rule[-0.200pt]{4.818pt}{0.400pt}}
\put(823.0,290.0){\rule[-0.200pt]{0.400pt}{15.418pt}}
\put(813.0,290.0){\rule[-0.200pt]{4.818pt}{0.400pt}}
\put(813.0,354.0){\rule[-0.200pt]{4.818pt}{0.400pt}}
\put(875.0,232.0){\rule[-0.200pt]{0.400pt}{19.031pt}}
\put(865.0,232.0){\rule[-0.200pt]{4.818pt}{0.400pt}}
\put(865.0,311.0){\rule[-0.200pt]{4.818pt}{0.400pt}}
\put(985.0,250.0){\rule[-0.200pt]{0.400pt}{20.717pt}}
\put(975.0,250.0){\rule[-0.200pt]{4.818pt}{0.400pt}}
\put(975.0,336.0){\rule[-0.200pt]{4.818pt}{0.400pt}}
\put(997.0,290.0){\rule[-0.200pt]{0.400pt}{24.090pt}}
\put(987.0,290.0){\rule[-0.200pt]{4.818pt}{0.400pt}}
\put(987.0,390.0){\rule[-0.200pt]{4.818pt}{0.400pt}}
\put(1102.0,135.0){\rule[-0.200pt]{0.400pt}{26.017pt}}
\put(1092.0,135.0){\rule[-0.200pt]{4.818pt}{0.400pt}}
\put(1092.0,243.0){\rule[-0.200pt]{4.818pt}{0.400pt}}
\put(1113.0,304.0){\rule[-0.200pt]{0.400pt}{27.703pt}}
\put(1103.0,304.0){\rule[-0.200pt]{4.818pt}{0.400pt}}
\put(1103.0,419.0){\rule[-0.200pt]{4.818pt}{0.400pt}}
\put(1398.0,175.0){\rule[-0.200pt]{0.400pt}{44.807pt}}
\put(1388.0,175.0){\rule[-0.200pt]{4.818pt}{0.400pt}}
\put(1388.0,361.0){\rule[-0.200pt]{4.818pt}{0.400pt}}
\end{picture}

%% file: b24_dgtd.tex
\setlength{\unitlength}{0.240900pt}
\ifx\plotpoint\undefined\newsavebox{\plotpoint}\fi
\begin{picture}(1500,900)(0,0)
\font\gnuplot=cmr10 at 12pt
\gnuplot
\sbox{\plotpoint}{\rule[-0.200pt]{0.400pt}{0.400pt}}%
\put(120.0,146.0){\rule[-0.200pt]{321.842pt}{0.400pt}}
\put(120.0,31.0){\rule[-0.200pt]{4.818pt}{0.400pt}}
\put(108,31){\makebox(0,0)[r]{$-0.02$}}
\put(1436.0,31.0){\rule[-0.200pt]{4.818pt}{0.400pt}}
\put(120.0,146.0){\rule[-0.200pt]{4.818pt}{0.400pt}}
\put(108,146){\makebox(0,0)[r]{$0$}}
\put(1436.0,146.0){\rule[-0.200pt]{4.818pt}{0.400pt}}
\put(120.0,261.0){\rule[-0.200pt]{4.818pt}{0.400pt}}
\put(108,261){\makebox(0,0)[r]{$0.02$}}
\put(1436.0,261.0){\rule[-0.200pt]{4.818pt}{0.400pt}}
\put(120.0,376.0){\rule[-0.200pt]{4.818pt}{0.400pt}}
\put(108,376){\makebox(0,0)[r]{$0.04$}}
\put(1436.0,376.0){\rule[-0.200pt]{4.818pt}{0.400pt}}
\put(120.0,491.0){\rule[-0.200pt]{4.818pt}{0.400pt}}
\put(108,491){\makebox(0,0)[r]{$0.06$}}
\put(1436.0,491.0){\rule[-0.200pt]{4.818pt}{0.400pt}}
\put(120.0,606.0){\rule[-0.200pt]{4.818pt}{0.400pt}}
\put(108,606){\makebox(0,0)[r]{$0.08$}}
\put(1436.0,606.0){\rule[-0.200pt]{4.818pt}{0.400pt}}
\put(120.0,721.0){\rule[-0.200pt]{4.818pt}{0.400pt}}
\put(108,721){\makebox(0,0)[r]{$0.1$}}
\put(1436.0,721.0){\rule[-0.200pt]{4.818pt}{0.400pt}}
\put(120.0,836.0){\rule[-0.200pt]{4.818pt}{0.400pt}}
\put(108,836){\makebox(0,0)[r]{$0.12$}}
\put(1436.0,836.0){\rule[-0.200pt]{4.818pt}{0.400pt}}
\put(199.0,31.0){\rule[-0.200pt]{0.400pt}{4.818pt}}
\put(199,19){\makebox(0,0){\shortstack{\\ \\ \\ \\ $5$}}}
\put(199.0,873.0){\rule[-0.200pt]{0.400pt}{4.818pt}}
\put(395.0,31.0){\rule[-0.200pt]{0.400pt}{4.818pt}}
\put(395,19){\makebox(0,0){\shortstack{\\ \\ \\ \\ $10$}}}
\put(395.0,873.0){\rule[-0.200pt]{0.400pt}{4.818pt}}
\put(592.0,31.0){\rule[-0.200pt]{0.400pt}{4.818pt}}
\put(592,19){\makebox(0,0){\shortstack{\\ \\ \\ \\ $15$}}}
\put(592.0,873.0){\rule[-0.200pt]{0.400pt}{4.818pt}}
\put(788.0,31.0){\rule[-0.200pt]{0.400pt}{4.818pt}}
\put(788,19){\makebox(0,0){\shortstack{\\ \\ \\ \\ $20$}}}
\put(788.0,873.0){\rule[-0.200pt]{0.400pt}{4.818pt}}
\put(984.0,31.0){\rule[-0.200pt]{0.400pt}{4.818pt}}
\put(984,19){\makebox(0,0){\shortstack{\\ \\ \\ \\ $25$}}}
\put(984.0,873.0){\rule[-0.200pt]{0.400pt}{4.818pt}}
\put(1181.0,31.0){\rule[-0.200pt]{0.400pt}{4.818pt}}
\put(1181,19){\makebox(0,0){\shortstack{\\ \\ \\ \\ $30$}}}
\put(1181.0,873.0){\rule[-0.200pt]{0.400pt}{4.818pt}}
\put(1377.0,31.0){\rule[-0.200pt]{0.400pt}{4.818pt}}
\put(1377,19){\makebox(0,0){\shortstack{\\ \\ \\ \\ $35$}}}
\put(1377.0,873.0){\rule[-0.200pt]{0.400pt}{4.818pt}}
\put(120.0,31.0){\rule[-0.200pt]{321.842pt}{0.400pt}}
\put(1456.0,31.0){\rule[-0.200pt]{0.400pt}{207.656pt}}
\put(120.0,893.0){\rule[-0.200pt]{321.842pt}{0.400pt}}
\put(12,438){\makebox(0,0){$\sigma_{eff} (r,t)$}}
\put(788,-53){\makebox(0,0){$\hbox{Area} = r \times t$}}
\put(120.0,31.0){\rule[-0.200pt]{0.400pt}{207.656pt}}
\put(159,537){\makebox(0,0){$+$}}
\put(234,537){\makebox(0,0){$+$}}
\put(242,537){\makebox(0,0){$+$}}
\put(313,537){\makebox(0,0){$+$}}
\put(320,542){\makebox(0,0){$+$}}
\put(356,519){\makebox(0,0){$+$}}
\put(391,542){\makebox(0,0){$+$}}
\put(399,542){\makebox(0,0){$+$}}
\put(470,514){\makebox(0,0){$+$}}
\put(478,519){\makebox(0,0){$+$}}
\put(588,519){\makebox(0,0){$+$}}
\put(595,531){\makebox(0,0){$+$}}
\put(631,525){\makebox(0,0){$+$}}
\put(705,531){\makebox(0,0){$+$}}
\put(713,525){\makebox(0,0){$+$}}
\put(784,531){\makebox(0,0){$+$}}
\put(792,519){\makebox(0,0){$+$}}
\put(984,502){\makebox(0,0){$+$}}
\put(1098,542){\makebox(0,0){$+$}}
\put(1106,537){\makebox(0,0){$+$}}
\put(1177,496){\makebox(0,0){$+$}}
\put(1185,554){\makebox(0,0){$+$}}
\put(1373,537){\makebox(0,0){$+$}}
\put(1381,473){\makebox(0,0){$+$}}
\put(1417,537){\makebox(0,0){$+$}}
\put(159.0,531.0){\rule[-0.200pt]{0.400pt}{2.650pt}}
\put(149.0,531.0){\rule[-0.200pt]{4.818pt}{0.400pt}}
\put(149.0,542.0){\rule[-0.200pt]{4.818pt}{0.400pt}}
\put(234.0,531.0){\rule[-0.200pt]{0.400pt}{2.650pt}}
\put(224.0,531.0){\rule[-0.200pt]{4.818pt}{0.400pt}}
\put(224.0,542.0){\rule[-0.200pt]{4.818pt}{0.400pt}}
\put(242.0,531.0){\rule[-0.200pt]{0.400pt}{2.650pt}}
\put(232.0,531.0){\rule[-0.200pt]{4.818pt}{0.400pt}}
\put(232.0,542.0){\rule[-0.200pt]{4.818pt}{0.400pt}}
\put(313.0,531.0){\rule[-0.200pt]{0.400pt}{2.650pt}}
\put(303.0,531.0){\rule[-0.200pt]{4.818pt}{0.400pt}}
\put(303.0,542.0){\rule[-0.200pt]{4.818pt}{0.400pt}}
\put(320.0,537.0){\rule[-0.200pt]{0.400pt}{2.650pt}}
\put(310.0,537.0){\rule[-0.200pt]{4.818pt}{0.400pt}}
\put(310.0,548.0){\rule[-0.200pt]{4.818pt}{0.400pt}}
\put(356.0,514.0){\rule[-0.200pt]{0.400pt}{2.650pt}}
\put(346.0,514.0){\rule[-0.200pt]{4.818pt}{0.400pt}}
\put(346.0,525.0){\rule[-0.200pt]{4.818pt}{0.400pt}}
\put(391.0,537.0){\rule[-0.200pt]{0.400pt}{2.650pt}}
\put(381.0,537.0){\rule[-0.200pt]{4.818pt}{0.400pt}}
\put(381.0,548.0){\rule[-0.200pt]{4.818pt}{0.400pt}}
\put(399.0,537.0){\rule[-0.200pt]{0.400pt}{2.650pt}}
\put(389.0,537.0){\rule[-0.200pt]{4.818pt}{0.400pt}}
\put(389.0,548.0){\rule[-0.200pt]{4.818pt}{0.400pt}}
\put(470.0,508.0){\rule[-0.200pt]{0.400pt}{2.650pt}}
\put(460.0,508.0){\rule[-0.200pt]{4.818pt}{0.400pt}}
\put(460.0,519.0){\rule[-0.200pt]{4.818pt}{0.400pt}}
\put(478.0,514.0){\rule[-0.200pt]{0.400pt}{2.650pt}}
\put(468.0,514.0){\rule[-0.200pt]{4.818pt}{0.400pt}}
\put(468.0,525.0){\rule[-0.200pt]{4.818pt}{0.400pt}}
\put(588.0,514.0){\rule[-0.200pt]{0.400pt}{2.650pt}}
\put(578.0,514.0){\rule[-0.200pt]{4.818pt}{0.400pt}}
\put(578.0,525.0){\rule[-0.200pt]{4.818pt}{0.400pt}}
\put(595.0,525.0){\rule[-0.200pt]{0.400pt}{2.891pt}}
\put(585.0,525.0){\rule[-0.200pt]{4.818pt}{0.400pt}}
\put(585.0,537.0){\rule[-0.200pt]{4.818pt}{0.400pt}}
\put(631.0,519.0){\rule[-0.200pt]{0.400pt}{2.891pt}}
\put(621.0,519.0){\rule[-0.200pt]{4.818pt}{0.400pt}}
\put(621.0,531.0){\rule[-0.200pt]{4.818pt}{0.400pt}}
\put(705.0,525.0){\rule[-0.200pt]{0.400pt}{2.891pt}}
\put(695.0,525.0){\rule[-0.200pt]{4.818pt}{0.400pt}}
\put(695.0,537.0){\rule[-0.200pt]{4.818pt}{0.400pt}}
\put(713.0,519.0){\rule[-0.200pt]{0.400pt}{2.891pt}}
\put(703.0,519.0){\rule[-0.200pt]{4.818pt}{0.400pt}}
\put(703.0,531.0){\rule[-0.200pt]{4.818pt}{0.400pt}}
\put(784.0,525.0){\rule[-0.200pt]{0.400pt}{2.891pt}}
\put(774.0,525.0){\rule[-0.200pt]{4.818pt}{0.400pt}}
\put(774.0,537.0){\rule[-0.200pt]{4.818pt}{0.400pt}}
\put(792.0,508.0){\rule[-0.200pt]{0.400pt}{5.541pt}}
\put(782.0,508.0){\rule[-0.200pt]{4.818pt}{0.400pt}}
\put(782.0,531.0){\rule[-0.200pt]{4.818pt}{0.400pt}}
\put(984.0,491.0){\rule[-0.200pt]{0.400pt}{5.541pt}}
\put(974.0,491.0){\rule[-0.200pt]{4.818pt}{0.400pt}}
\put(974.0,514.0){\rule[-0.200pt]{4.818pt}{0.400pt}}
\put(1098.0,525.0){\rule[-0.200pt]{0.400pt}{8.431pt}}
\put(1088.0,525.0){\rule[-0.200pt]{4.818pt}{0.400pt}}
\put(1088.0,560.0){\rule[-0.200pt]{4.818pt}{0.400pt}}
\put(1106.0,519.0){\rule[-0.200pt]{0.400pt}{8.431pt}}
\put(1096.0,519.0){\rule[-0.200pt]{4.818pt}{0.400pt}}
\put(1096.0,554.0){\rule[-0.200pt]{4.818pt}{0.400pt}}
\put(1177.0,479.0){\rule[-0.200pt]{0.400pt}{8.431pt}}
\put(1167.0,479.0){\rule[-0.200pt]{4.818pt}{0.400pt}}
\put(1167.0,514.0){\rule[-0.200pt]{4.818pt}{0.400pt}}
\put(1185.0,537.0){\rule[-0.200pt]{0.400pt}{8.191pt}}
\put(1175.0,537.0){\rule[-0.200pt]{4.818pt}{0.400pt}}
\put(1175.0,571.0){\rule[-0.200pt]{4.818pt}{0.400pt}}
\put(1373.0,514.0){\rule[-0.200pt]{0.400pt}{11.081pt}}
\put(1363.0,514.0){\rule[-0.200pt]{4.818pt}{0.400pt}}
\put(1363.0,560.0){\rule[-0.200pt]{4.818pt}{0.400pt}}
\put(1381.0,451.0){\rule[-0.200pt]{0.400pt}{10.840pt}}
\put(1371.0,451.0){\rule[-0.200pt]{4.818pt}{0.400pt}}
\put(1371.0,496.0){\rule[-0.200pt]{4.818pt}{0.400pt}}
\put(1417.0,508.0){\rule[-0.200pt]{0.400pt}{13.731pt}}
\put(1407.0,508.0){\rule[-0.200pt]{4.818pt}{0.400pt}}
\put(1407.0,565.0){\rule[-0.200pt]{4.818pt}{0.400pt}}
\put(159,870){\makebox(0,0){$\times$}}
\put(234,755){\makebox(0,0){$\times$}}
\put(242,755){\makebox(0,0){$\times$}}
\put(320,703){\makebox(0,0){$\times$}}
\put(391,686){\makebox(0,0){$\times$}}
\put(399,669){\makebox(0,0){$\times$}}
\put(470,514){\makebox(0,0){$\times$}}
\put(478,542){\makebox(0,0){$\times$}}
\put(588,525){\makebox(0,0){$\times$}}
\put(595,491){\makebox(0,0){$\times$}}
\put(631,491){\makebox(0,0){$\times$}}
\put(705,485){\makebox(0,0){$\times$}}
\put(713,462){\makebox(0,0){$\times$}}
\put(784,387){\makebox(0,0){$\times$}}
\put(792,439){\makebox(0,0){$\times$}}
\put(984,318){\makebox(0,0){$\times$}}
\put(1098,226){\makebox(0,0){$\times$}}
\put(1106,382){\makebox(0,0){$\times$}}
\put(1177,301){\makebox(0,0){$\times$}}
\put(1185,180){\makebox(0,0){$\times$}}
\put(1373,175){\makebox(0,0){$\times$}}
\put(1381,318){\makebox(0,0){$\times$}}
\put(1417,422){\makebox(0,0){$\times$}}
\put(159.0,859.0){\rule[-0.200pt]{0.400pt}{5.541pt}}
\put(149.0,859.0){\rule[-0.200pt]{4.818pt}{0.400pt}}
\put(149.0,882.0){\rule[-0.200pt]{4.818pt}{0.400pt}}
\put(234.0,744.0){\rule[-0.200pt]{0.400pt}{5.541pt}}
\put(224.0,744.0){\rule[-0.200pt]{4.818pt}{0.400pt}}
\put(224.0,767.0){\rule[-0.200pt]{4.818pt}{0.400pt}}
\put(242.0,744.0){\rule[-0.200pt]{0.400pt}{5.541pt}}
\put(232.0,744.0){\rule[-0.200pt]{4.818pt}{0.400pt}}
\put(232.0,767.0){\rule[-0.200pt]{4.818pt}{0.400pt}}
\put(320.0,692.0){\rule[-0.200pt]{0.400pt}{5.541pt}}
\put(310.0,692.0){\rule[-0.200pt]{4.818pt}{0.400pt}}
\put(310.0,715.0){\rule[-0.200pt]{4.818pt}{0.400pt}}
\put(391.0,669.0){\rule[-0.200pt]{0.400pt}{8.191pt}}
\put(381.0,669.0){\rule[-0.200pt]{4.818pt}{0.400pt}}
\put(381.0,703.0){\rule[-0.200pt]{4.818pt}{0.400pt}}
\put(399.0,652.0){\rule[-0.200pt]{0.400pt}{8.191pt}}
\put(389.0,652.0){\rule[-0.200pt]{4.818pt}{0.400pt}}
\put(389.0,686.0){\rule[-0.200pt]{4.818pt}{0.400pt}}
\put(470.0,496.0){\rule[-0.200pt]{0.400pt}{8.431pt}}
\put(460.0,496.0){\rule[-0.200pt]{4.818pt}{0.400pt}}
\put(460.0,531.0){\rule[-0.200pt]{4.818pt}{0.400pt}}
\put(478.0,525.0){\rule[-0.200pt]{0.400pt}{8.431pt}}
\put(468.0,525.0){\rule[-0.200pt]{4.818pt}{0.400pt}}
\put(468.0,560.0){\rule[-0.200pt]{4.818pt}{0.400pt}}
\put(588.0,502.0){\rule[-0.200pt]{0.400pt}{11.081pt}}
\put(578.0,502.0){\rule[-0.200pt]{4.818pt}{0.400pt}}
\put(578.0,548.0){\rule[-0.200pt]{4.818pt}{0.400pt}}
\put(595.0,468.0){\rule[-0.200pt]{0.400pt}{11.081pt}}
\put(585.0,468.0){\rule[-0.200pt]{4.818pt}{0.400pt}}
\put(585.0,514.0){\rule[-0.200pt]{4.818pt}{0.400pt}}
\put(631.0,468.0){\rule[-0.200pt]{0.400pt}{11.081pt}}
\put(621.0,468.0){\rule[-0.200pt]{4.818pt}{0.400pt}}
\put(621.0,514.0){\rule[-0.200pt]{4.818pt}{0.400pt}}
\put(705.0,456.0){\rule[-0.200pt]{0.400pt}{13.972pt}}
\put(695.0,456.0){\rule[-0.200pt]{4.818pt}{0.400pt}}
\put(695.0,514.0){\rule[-0.200pt]{4.818pt}{0.400pt}}
\put(713.0,439.0){\rule[-0.200pt]{0.400pt}{11.081pt}}
\put(703.0,439.0){\rule[-0.200pt]{4.818pt}{0.400pt}}
\put(703.0,485.0){\rule[-0.200pt]{4.818pt}{0.400pt}}
\put(784.0,347.0){\rule[-0.200pt]{0.400pt}{19.513pt}}
\put(774.0,347.0){\rule[-0.200pt]{4.818pt}{0.400pt}}
\put(774.0,428.0){\rule[-0.200pt]{4.818pt}{0.400pt}}
\put(792.0,399.0){\rule[-0.200pt]{0.400pt}{19.272pt}}
\put(782.0,399.0){\rule[-0.200pt]{4.818pt}{0.400pt}}
\put(782.0,479.0){\rule[-0.200pt]{4.818pt}{0.400pt}}
\put(984.0,255.0){\rule[-0.200pt]{0.400pt}{30.594pt}}
\put(974.0,255.0){\rule[-0.200pt]{4.818pt}{0.400pt}}
\put(974.0,382.0){\rule[-0.200pt]{4.818pt}{0.400pt}}
\put(1098.0,152.0){\rule[-0.200pt]{0.400pt}{35.894pt}}
\put(1088.0,152.0){\rule[-0.200pt]{4.818pt}{0.400pt}}
\put(1088.0,301.0){\rule[-0.200pt]{4.818pt}{0.400pt}}
\put(1106.0,295.0){\rule[-0.200pt]{0.400pt}{41.676pt}}
\put(1096.0,295.0){\rule[-0.200pt]{4.818pt}{0.400pt}}
\put(1096.0,468.0){\rule[-0.200pt]{4.818pt}{0.400pt}}
\put(1177.0,198.0){\rule[-0.200pt]{0.400pt}{49.866pt}}
\put(1167.0,198.0){\rule[-0.200pt]{4.818pt}{0.400pt}}
\put(1167.0,405.0){\rule[-0.200pt]{4.818pt}{0.400pt}}
\put(1185.0,94.0){\rule[-0.200pt]{0.400pt}{41.676pt}}
\put(1175.0,94.0){\rule[-0.200pt]{4.818pt}{0.400pt}}
\put(1175.0,267.0){\rule[-0.200pt]{4.818pt}{0.400pt}}
\put(1373.0,42.0){\rule[-0.200pt]{0.400pt}{63.838pt}}
\put(1363.0,42.0){\rule[-0.200pt]{4.818pt}{0.400pt}}
\put(1363.0,307.0){\rule[-0.200pt]{4.818pt}{0.400pt}}
\put(1381.0,180.0){\rule[-0.200pt]{0.400pt}{66.488pt}}
\put(1371.0,180.0){\rule[-0.200pt]{4.818pt}{0.400pt}}
\put(1371.0,456.0){\rule[-0.200pt]{4.818pt}{0.400pt}}
\put(1417.0,307.0){\rule[-0.200pt]{0.400pt}{55.407pt}}
\put(1407.0,307.0){\rule[-0.200pt]{4.818pt}{0.400pt}}
\put(1407.0,537.0){\rule[-0.200pt]{4.818pt}{0.400pt}}
\end{picture}

%% file: b23_ddgt.tex
\setlength{\unitlength}{0.240900pt}
\ifx\plotpoint\undefined\newsavebox{\plotpoint}\fi
\begin{picture}(1500,900)(0,0)
\font\gnuplot=cmr10 at 12pt
\gnuplot
\sbox{\plotpoint}{\rule[-0.200pt]{0.400pt}{0.400pt}}%
\put(120.0,103.0){\rule[-0.200pt]{321.842pt}{0.400pt}}
\put(120.0,103.0){\rule[-0.200pt]{4.818pt}{0.400pt}}
\put(108,103){\makebox(0,0)[r]{$0$}}
\put(1436.0,103.0){\rule[-0.200pt]{4.818pt}{0.400pt}}
\put(120.0,282.0){\rule[-0.200pt]{4.818pt}{0.400pt}}
\put(108,282){\makebox(0,0)[r]{$0.05$}}
\put(1436.0,282.0){\rule[-0.200pt]{4.818pt}{0.400pt}}
\put(120.0,462.0){\rule[-0.200pt]{4.818pt}{0.400pt}}
\put(108,462){\makebox(0,0)[r]{$0.1$}}
\put(1436.0,462.0){\rule[-0.200pt]{4.818pt}{0.400pt}}
\put(120.0,642.0){\rule[-0.200pt]{4.818pt}{0.400pt}}
\put(108,642){\makebox(0,0)[r]{$0.15$}}
\put(1436.0,642.0){\rule[-0.200pt]{4.818pt}{0.400pt}}
\put(120.0,821.0){\rule[-0.200pt]{4.818pt}{0.400pt}}
\put(108,821){\makebox(0,0)[r]{$0.2$}}
\put(1436.0,821.0){\rule[-0.200pt]{4.818pt}{0.400pt}}
\put(236.0,31.0){\rule[-0.200pt]{0.400pt}{4.818pt}}
\put(236,19){\makebox(0,0){\shortstack{\\ \\ \\ \\ $5$}}}
\put(236.0,873.0){\rule[-0.200pt]{0.400pt}{4.818pt}}
\put(527.0,31.0){\rule[-0.200pt]{0.400pt}{4.818pt}}
\put(527,19){\makebox(0,0){\shortstack{\\ \\ \\ \\ $10$}}}
\put(527.0,873.0){\rule[-0.200pt]{0.400pt}{4.818pt}}
\put(817.0,31.0){\rule[-0.200pt]{0.400pt}{4.818pt}}
\put(817,19){\makebox(0,0){\shortstack{\\ \\ \\ \\ $15$}}}
\put(817.0,873.0){\rule[-0.200pt]{0.400pt}{4.818pt}}
\put(1107.0,31.0){\rule[-0.200pt]{0.400pt}{4.818pt}}
\put(1107,19){\makebox(0,0){\shortstack{\\ \\ \\ \\ $20$}}}
\put(1107.0,873.0){\rule[-0.200pt]{0.400pt}{4.818pt}}
\put(1398.0,31.0){\rule[-0.200pt]{0.400pt}{4.818pt}}
\put(1398,19){\makebox(0,0){\shortstack{\\ \\ \\ \\ $25$}}}
\put(1398.0,873.0){\rule[-0.200pt]{0.400pt}{4.818pt}}
\put(120.0,31.0){\rule[-0.200pt]{321.842pt}{0.400pt}}
\put(1456.0,31.0){\rule[-0.200pt]{0.400pt}{207.656pt}}
\put(120.0,893.0){\rule[-0.200pt]{321.842pt}{0.400pt}}
\put(12,534){\makebox(0,0){$\sigma_{\hbox{\small eff}} (r,t)$}}
\put(788,-53){\makebox(0,0){$\hbox{Area} = r \times t$}}
\put(120.0,31.0){\rule[-0.200pt]{0.400pt}{207.656pt}}
\put(178,846){\makebox(0,0){$\times$}}
\put(288,807){\makebox(0,0){$\times$}}
\put(300,810){\makebox(0,0){$\times$}}
\put(405,764){\makebox(0,0){$\times$}}
\put(416,782){\makebox(0,0){$\times$}}
\put(469,681){\makebox(0,0){$\times$}}
\put(521,771){\makebox(0,0){$\times$}}
\put(532,757){\makebox(0,0){$\times$}}
\put(637,642){\makebox(0,0){$\times$}}
\put(649,620){\makebox(0,0){$\times$}}
\put(811,627){\makebox(0,0){$\times$}}
\put(823,577){\makebox(0,0){$\times$}}
\put(875,502){\makebox(0,0){$\times$}}
\put(985,717){\makebox(0,0){$\times$}}
\put(997,379){\makebox(0,0){$\times$}}
\put(1102,379){\makebox(0,0){$\times$}}
\put(1113,275){\makebox(0,0){$\times$}}
\put(1398,588){\makebox(0,0){$\times$}}
\put(178.0,839.0){\rule[-0.200pt]{0.400pt}{3.373pt}}
\put(168.0,839.0){\rule[-0.200pt]{4.818pt}{0.400pt}}
\put(168.0,853.0){\rule[-0.200pt]{4.818pt}{0.400pt}}
\put(288.0,796.0){\rule[-0.200pt]{0.400pt}{5.300pt}}
\put(278.0,796.0){\rule[-0.200pt]{4.818pt}{0.400pt}}
\put(278.0,818.0){\rule[-0.200pt]{4.818pt}{0.400pt}}
\put(300.0,800.0){\rule[-0.200pt]{0.400pt}{5.059pt}}
\put(290.0,800.0){\rule[-0.200pt]{4.818pt}{0.400pt}}
\put(290.0,821.0){\rule[-0.200pt]{4.818pt}{0.400pt}}
\put(405.0,749.0){\rule[-0.200pt]{0.400pt}{6.986pt}}
\put(395.0,749.0){\rule[-0.200pt]{4.818pt}{0.400pt}}
\put(395.0,778.0){\rule[-0.200pt]{4.818pt}{0.400pt}}
\put(416.0,767.0){\rule[-0.200pt]{0.400pt}{6.986pt}}
\put(406.0,767.0){\rule[-0.200pt]{4.818pt}{0.400pt}}
\put(406.0,796.0){\rule[-0.200pt]{4.818pt}{0.400pt}}
\put(469.0,667.0){\rule[-0.200pt]{0.400pt}{6.745pt}}
\put(459.0,667.0){\rule[-0.200pt]{4.818pt}{0.400pt}}
\put(459.0,695.0){\rule[-0.200pt]{4.818pt}{0.400pt}}
\put(521.0,753.0){\rule[-0.200pt]{0.400pt}{8.672pt}}
\put(511.0,753.0){\rule[-0.200pt]{4.818pt}{0.400pt}}
\put(511.0,789.0){\rule[-0.200pt]{4.818pt}{0.400pt}}
\put(532.0,742.0){\rule[-0.200pt]{0.400pt}{6.986pt}}
\put(522.0,742.0){\rule[-0.200pt]{4.818pt}{0.400pt}}
\put(522.0,771.0){\rule[-0.200pt]{4.818pt}{0.400pt}}
\put(637.0,616.0){\rule[-0.200pt]{0.400pt}{12.286pt}}
\put(627.0,616.0){\rule[-0.200pt]{4.818pt}{0.400pt}}
\put(627.0,667.0){\rule[-0.200pt]{4.818pt}{0.400pt}}
\put(649.0,591.0){\rule[-0.200pt]{0.400pt}{13.972pt}}
\put(639.0,591.0){\rule[-0.200pt]{4.818pt}{0.400pt}}
\put(639.0,649.0){\rule[-0.200pt]{4.818pt}{0.400pt}}
\put(811.0,581.0){\rule[-0.200pt]{0.400pt}{22.404pt}}
\put(801.0,581.0){\rule[-0.200pt]{4.818pt}{0.400pt}}
\put(801.0,674.0){\rule[-0.200pt]{4.818pt}{0.400pt}}
\put(823.0,523.0){\rule[-0.200pt]{0.400pt}{26.017pt}}
\put(813.0,523.0){\rule[-0.200pt]{4.818pt}{0.400pt}}
\put(813.0,631.0){\rule[-0.200pt]{4.818pt}{0.400pt}}
\put(875.0,458.0){\rule[-0.200pt]{0.400pt}{20.958pt}}
\put(865.0,458.0){\rule[-0.200pt]{4.818pt}{0.400pt}}
\put(865.0,545.0){\rule[-0.200pt]{4.818pt}{0.400pt}}
\put(985.0,656.0){\rule[-0.200pt]{0.400pt}{29.390pt}}
\put(975.0,656.0){\rule[-0.200pt]{4.818pt}{0.400pt}}
\put(975.0,778.0){\rule[-0.200pt]{4.818pt}{0.400pt}}
\put(997.0,315.0){\rule[-0.200pt]{0.400pt}{31.076pt}}
\put(987.0,315.0){\rule[-0.200pt]{4.818pt}{0.400pt}}
\put(987.0,444.0){\rule[-0.200pt]{4.818pt}{0.400pt}}
\put(1102.0,293.0){\rule[-0.200pt]{0.400pt}{41.676pt}}
\put(1092.0,293.0){\rule[-0.200pt]{4.818pt}{0.400pt}}
\put(1092.0,466.0){\rule[-0.200pt]{4.818pt}{0.400pt}}
\put(1113.0,185.0){\rule[-0.200pt]{0.400pt}{43.362pt}}
\put(1103.0,185.0){\rule[-0.200pt]{4.818pt}{0.400pt}}
\put(1103.0,365.0){\rule[-0.200pt]{4.818pt}{0.400pt}}
\put(1398.0,376.0){\rule[-0.200pt]{0.400pt}{102.142pt}}
\put(1388.0,376.0){\rule[-0.200pt]{4.818pt}{0.400pt}}
\put(1388.0,800.0){\rule[-0.200pt]{4.818pt}{0.400pt}}
\end{picture}

%% file: b24_ddgt.tex
\setlength{\unitlength}{0.240900pt}
\ifx\plotpoint\undefined\newsavebox{\plotpoint}\fi
\begin{picture}(1500,900)(0,0)
\font\gnuplot=cmr10 at 12pt
\gnuplot
\sbox{\plotpoint}{\rule[-0.200pt]{0.400pt}{0.400pt}}%
\put(120.0,146.0){\rule[-0.200pt]{321.842pt}{0.400pt}}
\put(120.0,31.0){\rule[-0.200pt]{4.818pt}{0.400pt}}
\put(108,31){\makebox(0,0)[r]{$-0.02$}}
\put(1436.0,31.0){\rule[-0.200pt]{4.818pt}{0.400pt}}
\put(120.0,146.0){\rule[-0.200pt]{4.818pt}{0.400pt}}
\put(108,146){\makebox(0,0)[r]{$0$}}
\put(1436.0,146.0){\rule[-0.200pt]{4.818pt}{0.400pt}}
\put(120.0,261.0){\rule[-0.200pt]{4.818pt}{0.400pt}}
\put(108,261){\makebox(0,0)[r]{$0.02$}}
\put(1436.0,261.0){\rule[-0.200pt]{4.818pt}{0.400pt}}
\put(120.0,376.0){\rule[-0.200pt]{4.818pt}{0.400pt}}
\put(108,376){\makebox(0,0)[r]{$0.04$}}
\put(1436.0,376.0){\rule[-0.200pt]{4.818pt}{0.400pt}}
\put(120.0,491.0){\rule[-0.200pt]{4.818pt}{0.400pt}}
\put(108,491){\makebox(0,0)[r]{$0.06$}}
\put(1436.0,491.0){\rule[-0.200pt]{4.818pt}{0.400pt}}
\put(120.0,606.0){\rule[-0.200pt]{4.818pt}{0.400pt}}
\put(108,606){\makebox(0,0)[r]{$0.08$}}
\put(1436.0,606.0){\rule[-0.200pt]{4.818pt}{0.400pt}}
\put(120.0,721.0){\rule[-0.200pt]{4.818pt}{0.400pt}}
\put(108,721){\makebox(0,0)[r]{$0.1$}}
\put(1436.0,721.0){\rule[-0.200pt]{4.818pt}{0.400pt}}
\put(120.0,836.0){\rule[-0.200pt]{4.818pt}{0.400pt}}
\put(108,836){\makebox(0,0)[r]{$0.12$}}
\put(1436.0,836.0){\rule[-0.200pt]{4.818pt}{0.400pt}}
\put(199.0,31.0){\rule[-0.200pt]{0.400pt}{4.818pt}}
\put(199,19){\makebox(0,0){\shortstack{\\ \\ \\ \\ $5$}}}
\put(199.0,873.0){\rule[-0.200pt]{0.400pt}{4.818pt}}
\put(395.0,31.0){\rule[-0.200pt]{0.400pt}{4.818pt}}
\put(395,19){\makebox(0,0){\shortstack{\\ \\ \\ \\ $10$}}}
\put(395.0,873.0){\rule[-0.200pt]{0.400pt}{4.818pt}}
\put(592.0,31.0){\rule[-0.200pt]{0.400pt}{4.818pt}}
\put(592,19){\makebox(0,0){\shortstack{\\ \\ \\ \\ $15$}}}
\put(592.0,873.0){\rule[-0.200pt]{0.400pt}{4.818pt}}
\put(788.0,31.0){\rule[-0.200pt]{0.400pt}{4.818pt}}
\put(788,19){\makebox(0,0){\shortstack{\\ \\ \\ \\ $20$}}}
\put(788.0,873.0){\rule[-0.200pt]{0.400pt}{4.818pt}}
\put(984.0,31.0){\rule[-0.200pt]{0.400pt}{4.818pt}}
\put(984,19){\makebox(0,0){\shortstack{\\ \\ \\ \\ $25$}}}
\put(984.0,873.0){\rule[-0.200pt]{0.400pt}{4.818pt}}
\put(1181.0,31.0){\rule[-0.200pt]{0.400pt}{4.818pt}}
\put(1181,19){\makebox(0,0){\shortstack{\\ \\ \\ \\ $30$}}}
\put(1181.0,873.0){\rule[-0.200pt]{0.400pt}{4.818pt}}
\put(1377.0,31.0){\rule[-0.200pt]{0.400pt}{4.818pt}}
\put(1377,19){\makebox(0,0){\shortstack{\\ \\ \\ \\ $35$}}}
\put(1377.0,873.0){\rule[-0.200pt]{0.400pt}{4.818pt}}
\put(120.0,31.0){\rule[-0.200pt]{321.842pt}{0.400pt}}
\put(1456.0,31.0){\rule[-0.200pt]{0.400pt}{207.656pt}}
\put(120.0,893.0){\rule[-0.200pt]{321.842pt}{0.400pt}}
\put(12,438){\makebox(0,0){$\sigma_{eff} (r,t)$}}
\put(788,-53){\makebox(0,0){$\hbox{Area} = r \times t$}}
\put(120.0,31.0){\rule[-0.200pt]{0.400pt}{207.656pt}}
\put(159,795){\makebox(0,0){$\times$}}
\put(234,767){\makebox(0,0){$\times$}}
\put(242,784){\makebox(0,0){$\times$}}
\put(313,761){\makebox(0,0){$\times$}}
\put(320,767){\makebox(0,0){$\times$}}
\put(356,721){\makebox(0,0){$\times$}}
\put(391,767){\makebox(0,0){$\times$}}
\put(399,778){\makebox(0,0){$\times$}}
\put(470,703){\makebox(0,0){$\times$}}
\put(478,698){\makebox(0,0){$\times$}}
\put(588,721){\makebox(0,0){$\times$}}
\put(595,703){\makebox(0,0){$\times$}}
\put(631,675){\makebox(0,0){$\times$}}
\put(705,675){\makebox(0,0){$\times$}}
\put(713,663){\makebox(0,0){$\times$}}
\put(784,634){\makebox(0,0){$\times$}}
\put(792,606){\makebox(0,0){$\times$}}
\put(984,508){\makebox(0,0){$\times$}}
\put(1098,594){\makebox(0,0){$\times$}}
\put(1106,479){\makebox(0,0){$\times$}}
\put(1177,468){\makebox(0,0){$\times$}}
\put(1185,462){\makebox(0,0){$\times$}}
\put(1373,577){\makebox(0,0){$\times$}}
\put(1381,376){\makebox(0,0){$\times$}}
\put(1417,410){\makebox(0,0){$\times$}}
\put(159.0,784.0){\rule[-0.200pt]{0.400pt}{5.541pt}}
\put(149.0,784.0){\rule[-0.200pt]{4.818pt}{0.400pt}}
\put(149.0,807.0){\rule[-0.200pt]{4.818pt}{0.400pt}}
\put(234.0,755.0){\rule[-0.200pt]{0.400pt}{5.541pt}}
\put(224.0,755.0){\rule[-0.200pt]{4.818pt}{0.400pt}}
\put(224.0,778.0){\rule[-0.200pt]{4.818pt}{0.400pt}}
\put(242.0,772.0){\rule[-0.200pt]{0.400pt}{5.541pt}}
\put(232.0,772.0){\rule[-0.200pt]{4.818pt}{0.400pt}}
\put(232.0,795.0){\rule[-0.200pt]{4.818pt}{0.400pt}}
\put(313.0,749.0){\rule[-0.200pt]{0.400pt}{5.541pt}}
\put(303.0,749.0){\rule[-0.200pt]{4.818pt}{0.400pt}}
\put(303.0,772.0){\rule[-0.200pt]{4.818pt}{0.400pt}}
\put(320.0,755.0){\rule[-0.200pt]{0.400pt}{5.541pt}}
\put(310.0,755.0){\rule[-0.200pt]{4.818pt}{0.400pt}}
\put(310.0,778.0){\rule[-0.200pt]{4.818pt}{0.400pt}}
\put(356.0,709.0){\rule[-0.200pt]{0.400pt}{5.541pt}}
\put(346.0,709.0){\rule[-0.200pt]{4.818pt}{0.400pt}}
\put(346.0,732.0){\rule[-0.200pt]{4.818pt}{0.400pt}}
\put(391.0,755.0){\rule[-0.200pt]{0.400pt}{5.541pt}}
\put(381.0,755.0){\rule[-0.200pt]{4.818pt}{0.400pt}}
\put(381.0,778.0){\rule[-0.200pt]{4.818pt}{0.400pt}}
\put(399.0,761.0){\rule[-0.200pt]{0.400pt}{8.191pt}}
\put(389.0,761.0){\rule[-0.200pt]{4.818pt}{0.400pt}}
\put(389.0,795.0){\rule[-0.200pt]{4.818pt}{0.400pt}}
\put(470.0,686.0){\rule[-0.200pt]{0.400pt}{8.431pt}}
\put(460.0,686.0){\rule[-0.200pt]{4.818pt}{0.400pt}}
\put(460.0,721.0){\rule[-0.200pt]{4.818pt}{0.400pt}}
\put(478.0,686.0){\rule[-0.200pt]{0.400pt}{5.541pt}}
\put(468.0,686.0){\rule[-0.200pt]{4.818pt}{0.400pt}}
\put(468.0,709.0){\rule[-0.200pt]{4.818pt}{0.400pt}}
\put(588.0,698.0){\rule[-0.200pt]{0.400pt}{11.081pt}}
\put(578.0,698.0){\rule[-0.200pt]{4.818pt}{0.400pt}}
\put(578.0,744.0){\rule[-0.200pt]{4.818pt}{0.400pt}}
\put(595.0,680.0){\rule[-0.200pt]{0.400pt}{11.081pt}}
\put(585.0,680.0){\rule[-0.200pt]{4.818pt}{0.400pt}}
\put(585.0,726.0){\rule[-0.200pt]{4.818pt}{0.400pt}}
\put(631.0,646.0){\rule[-0.200pt]{0.400pt}{13.731pt}}
\put(621.0,646.0){\rule[-0.200pt]{4.818pt}{0.400pt}}
\put(621.0,703.0){\rule[-0.200pt]{4.818pt}{0.400pt}}
\put(705.0,646.0){\rule[-0.200pt]{0.400pt}{13.731pt}}
\put(695.0,646.0){\rule[-0.200pt]{4.818pt}{0.400pt}}
\put(695.0,703.0){\rule[-0.200pt]{4.818pt}{0.400pt}}
\put(713.0,640.0){\rule[-0.200pt]{0.400pt}{11.081pt}}
\put(703.0,640.0){\rule[-0.200pt]{4.818pt}{0.400pt}}
\put(703.0,686.0){\rule[-0.200pt]{4.818pt}{0.400pt}}
\put(784.0,600.0){\rule[-0.200pt]{0.400pt}{16.622pt}}
\put(774.0,600.0){\rule[-0.200pt]{4.818pt}{0.400pt}}
\put(774.0,669.0){\rule[-0.200pt]{4.818pt}{0.400pt}}
\put(792.0,560.0){\rule[-0.200pt]{0.400pt}{22.163pt}}
\put(782.0,560.0){\rule[-0.200pt]{4.818pt}{0.400pt}}
\put(782.0,652.0){\rule[-0.200pt]{4.818pt}{0.400pt}}
\put(984.0,451.0){\rule[-0.200pt]{0.400pt}{27.463pt}}
\put(974.0,451.0){\rule[-0.200pt]{4.818pt}{0.400pt}}
\put(974.0,565.0){\rule[-0.200pt]{4.818pt}{0.400pt}}
\put(1098.0,508.0){\rule[-0.200pt]{0.400pt}{41.435pt}}
\put(1088.0,508.0){\rule[-0.200pt]{4.818pt}{0.400pt}}
\put(1088.0,680.0){\rule[-0.200pt]{4.818pt}{0.400pt}}
\put(1106.0,382.0){\rule[-0.200pt]{0.400pt}{46.975pt}}
\put(1096.0,382.0){\rule[-0.200pt]{4.818pt}{0.400pt}}
\put(1096.0,577.0){\rule[-0.200pt]{4.818pt}{0.400pt}}
\put(1177.0,382.0){\rule[-0.200pt]{0.400pt}{41.435pt}}
\put(1167.0,382.0){\rule[-0.200pt]{4.818pt}{0.400pt}}
\put(1167.0,554.0){\rule[-0.200pt]{4.818pt}{0.400pt}}
\put(1185.0,393.0){\rule[-0.200pt]{0.400pt}{33.244pt}}
\put(1175.0,393.0){\rule[-0.200pt]{4.818pt}{0.400pt}}
\put(1175.0,531.0){\rule[-0.200pt]{4.818pt}{0.400pt}}
\put(1373.0,433.0){\rule[-0.200pt]{0.400pt}{69.379pt}}
\put(1363.0,433.0){\rule[-0.200pt]{4.818pt}{0.400pt}}
\put(1363.0,721.0){\rule[-0.200pt]{4.818pt}{0.400pt}}
\put(1381.0,278.0){\rule[-0.200pt]{0.400pt}{46.975pt}}
\put(1371.0,278.0){\rule[-0.200pt]{4.818pt}{0.400pt}}
\put(1371.0,473.0){\rule[-0.200pt]{4.818pt}{0.400pt}}
\put(1417.0,267.0){\rule[-0.200pt]{0.400pt}{69.138pt}}
\put(1407.0,267.0){\rule[-0.200pt]{4.818pt}{0.400pt}}
\put(1407.0,554.0){\rule[-0.200pt]{4.818pt}{0.400pt}}
\end{picture}

%% file: mag_gribov.bbl
\begin{thebibliography}{99}

\bibitem{tHo}
G. 't Hooft ---
{\sl Nucl. Phys.} {\bf B190} (1981) 455.

\bibitem{suzuki}
T. Suzuki and I. Yotsuyanagi ---
{\sl Phys. Rev. D} {\bf 42} (1990) 4257.

\bibitem{MAG}
A.S. Kronfeld, M.L. Laursen, G. Schierholz and U.-J. Wiese ---
{\sl Phys. Lett.} {\bf 198B} (1987) 516.

\bibitem{magrev}
T. Suzuki ---
{\sl Nucl. Phys.} {\bf B30} (Proc.Suppl.) (1993) 176.

\bibitem{stack2}
J. Stack and R. Wensley ---
{\sl Phys. Rev. D} {\bf 50} (1994) 3399.

\bibitem{suzuki2}
S. Hioki, S. Kitahara, Y. Matsubara, O. Miyamura, S. Ohno and
T. Suzuki ---
{\sl Phys. Lett.} {\bf 271B} (1991) 201.

\bibitem{bali_Rmax}
G. Bali, V. Bornyakov, M. M\"uller-Preussker and F. Pahl ---
{\sl Nucl. Phys.} {\bf B42} (Proc.Suppl.) (1995) 852.

\bibitem{tep3}
A. Hart and M. Teper  ---
in progress.

\bibitem{DeGT}
T.A. DeGrand and D. Toussaint ---
{\sl Phys. Rev. D} {\bf 22} (1980) 2478.

\bibitem{stackvecpot}
J. Stack and R. Wensley ---
{\sl Nucl. Phys.} {\bf B371} (1992) 597.

\bibitem{suzukipoly}
T. Suzuki ---
Preprint {\sl  KANAZAWA-95-02} available as {\sl hep--lat/9506016}.

\bibitem{michtep}
C. Michael and M. Teper ---
{\sl Nucl. Phys.} {\bf B305} (1988) 453.

\bibitem{zwanz}
D. Zwanziger ---
{\sl Nucl. Phys.} {\bf B412} (1994) 657.

\bibitem{Mitr1}
V. Bornyakov, V. Mitrjushkin, M. M\"uller-Preussker and F. Pahl ---
{\sl Phys. Lett.} {\bf B317} (1993) 596.

\bibitem{Mitr2}
V. Mitrjushkin --- private communication

\bibitem{Polybook}
A.M. Polyakov ---
{\sl Nucl. Phys.} {\bf B120} (1977) 429.

\bibitem{tep}
M. Teper ---
{\sl Phys. Lett.} {\bf B311} (1993) 223.

\bibitem{Balinew}
G. Bali, V. Bornyakov, M. M\"uller-Preussker and K. Schilling ---
Preprint {\sl SHEP 95/42} available as {\sl hep--lat/9603012}.

\end{thebibliography}
